\documentclass[preprintnumbers,amsmath,amssymb,floatfix,10pt,prd,superscriptaddress,nofootinbib,twocolumn]{revtex4}

\usepackage{silence}
\usepackage{doi}
\usepackage{hyperref}
\hypersetup{
  colorlinks=true,        
  linkcolor=magenta,         
  citecolor=red,         
}

\usepackage{graphicx}
\usepackage{dcolumn}
\usepackage{bm}
\usepackage{booktabs}
\usepackage{multirow}
\usepackage{color}
\usepackage{enumitem}
\usepackage{amsmath}
\usepackage{amssymb}

\newcommand{\beq}{\begin{equation}}
\newcommand{\eeq}{\end{equation}}
\newcommand{\bea}{\begin{eqnarray}}
\newcommand{\eea}{\end{eqnarray}}

\usepackage{bbm}
\usepackage{amsfonts}
\usepackage{mathrsfs}
\usepackage{latexsym}
\usepackage{epsfig}
\usepackage{epstopdf}
\usepackage{epstopdf}
\usepackage{graphicx}
\usepackage{amssymb}
\usepackage{amsmath}
\usepackage{dcolumn}
\usepackage{bm}
\usepackage{color}
\usepackage{comment}
\usepackage{xcolor}
\usepackage{subfigure}
\usepackage{orcidlink}

\begin{document}
\title{Lyapunov Stability of a Schwarzschild Black Hole Surrounded by the Kiselev Anisotropic Fluid}
\author{Saswati Roy\, \orcidlink{0000-0002-7028-2627}}
\email{sr.phy2011@yahoo.com}
\affiliation{Department of Physics, National Institute of Technology, Agartala, Tripura--799046, India}

\author{Paramita Debnath\, \orcidlink{}}
\email{paramitad777@gmail.com}
\affiliation{Department of Physics, National Institute of Technology, Agartala, Tripura--799046, India}

\author{Shubham Kala\orcidlink{0000-0003-2379-0204}}
\email{shubhamkala871@gmail.com}
\affiliation{Centre for Space Research, North-West University, Potchefstroom 2520,
South Africa}

\author{Hemwati Nandan\, \orcidlink{0000-0002-1183-4727}}
\email{hnandan@associates.iucaa.in}
\affiliation{Department of Physics, Hemwati Nandan Bahuguna Garhwal University,
Srinagar--246174, Uttarakhand, India}
\affiliation{Centre for Space Research, North-West University,
Potchefstroom 2520, South Africa}

\begin{abstract}
\noindent
We investigate the geodesic structure and the stability of circular orbits around a Schwarzschild black hole surrounded by a Kiselev anisotropic fluid, characterised by the equation-of-state parameter $\omega$ and the normalisation parameter $N$. We consider six representative values, $\omega=-1,-2/3,-1/3,0,1/3$ and $1$, over the admissible range of $N$ determined by the horizon structure, including the corresponding extremal limits. Using the effective potentials for null and timelike geodesics, we determine the photon sphere, the innermost stable circular orbit (ISCO) and the outermost stable circular orbit (OSCO), and examine their stability through coordinate-time Lyapunov exponents. The photon sphere is unstable for all values of $\omega$ and $N$ considered. Its instability weakens with increasing $N$ for $\omega\leq0$ and vanishes at extremality for $\omega=-1$ and $-2/3$, whereas it varies non-monotonically with $N$ for $\omega=1/3$ and strengthens for $\omega=1$. For $\omega=-1$ and $-2/3$, stable circular orbits are confined between the ISCO and the OSCO, which merge at a critical value of $N$ beyond which no stable circular orbits exist. In contrast, for $\omega\geq-1/3$ stable circular orbits exist at all radii beyond the ISCO, whose location varies with $N$. We also derive a closed-form expression for the squared coordinate-time Lyapunov exponent $(\lambda_t^T)^2$, valid for arbitrary $\omega$, whose sign changes exactly at the ISCO and OSCO, so that the Lyapunov criterion reproduces the boundaries of stable circular motion. The Schwarzschild--de Sitter, mass-shifted Schwarzschild and Reissner--Nordstr\"om geometries are recovered as special cases, and the results may have implications for black-hole shadows, thin accretion discs and quasi-periodic oscillations.\\


\end{abstract}

\maketitle

\newpage
\section{Introduction}
\label{sec:intro}
The discovery of accelerated expansion of the Universe from observations of distant Type Ia supernovae made dark energy one of the central components of modern cosmology~\cite{SupernovaSearchTeam:1998fmf,SupernovaCosmologyProject:1998vns}. Measurements of the cosmic microwave background have since supported this picture and constrained the cosmological parameters within the $\Lambda$CDM framework~\cite{Planck:2018vyg}. Although the cosmological constant $\Lambda$ offers the simplest explanation of the acceleration, the unexplained origin of its value, together with the associated fine-tuning and coincidence problems, has motivated alternative models of dark energy. Among these, quintessence describes dark energy as a dynamical component with an equation-of-state~(EoS) parameter in the range $-1<\omega<-1/3$~\cite{Copeland:2006wr,tsujikawa2013quintessence}. Unlike $\Lambda$, whose energy density is constant, quintessence allows the dark-energy density to evolve as the Universe expands. This has also motivated studies of how dark-energy-like matter distributions affect the gravitational field around compact astrophysical objects.

Astrophysical black holes~(BHs) are not isolated in an ideal vacuum, but are generally surrounded by matter and energy~\cite{Kimball:2020opk,Figueiredo:2023gas,Johnson:2024ttr,Rahman:2023sof}. Understanding how this environment modifies the spacetime and the motion of test particles is therefore important for interpreting their observational signatures. Kiselev obtained a static, spherically symmetric BH solution surrounded by a matter distribution characterised by an EoS parameter $\omega$ and a normalisation parameter $N$~\cite{kiselev2003quintessence}. Its metric function contains an additional term proportional to $N/r^{3\omega+1}$, which modifies the spacetime geometry and hence the trajectories of test particles and photons. The Kiselev source is, however, more appropriately interpreted as an anisotropic fluid than as a canonical quintessence scalar field. As discussed by Visser, $\omega$ is the ratio of the angle-averaged pressure to the energy density of this anisotropic distribution~\cite{Visser:2019brz}. Depending on its value, the surrounding matter ranges from a component equivalent to a cosmological constant to radiation like and stiff matter like distributions. The Kiselev solution therefore provides a useful framework for examining how different surrounding matter distributions influence BH spacetimes.

The motion of particles and photons on circular geodesics provides an important link between the geometry of a BH spacetime and its astrophysical observables~\cite{Young:1976zz,Levin:2008mq,Pugliese:2010ps,Pradhan:2010ws,Pradhan:2014zia,Tursunov:2016dss,Singh:2026odr,Singh:2026vfd}. The unstable circular photon orbit determines the critical impact parameter associated with the BH shadow, which has become directly relevant since the Event Horizon Telescope observations of the supermassive BHs M87* and Sgr A*~\cite{EventHorizonTelescope:2019dse}. For massive particles, the innermost stable circular orbit (ISCO) marks the inner edge of a geometrically thin accretion disc in standard disc models and largely determines the radiative efficiency of accretion~\cite{Bardeen:1972fi,Suzuki:1997by}. When an outermost stable circular orbit (OSCO) also exists, the region of stable circular motion is bounded on both sides. Circular geodesics also underlie several models of the quasi-periodic oscillations observed in accreting compact objects~\cite{PhysRevD.103.104070}. Since $N$ and $\omega$ enter the effective potential directly, they can shift the locations of circular orbits and alter their stability and may therefore leave observable imprints on BH shadows, accretion processes, and orbital frequencies.

Several aspects of geodesic motion around the Kiselev BHs have already been studied. Fernando investigated null geodesics and the corresponding optical properties of these spacetimes~\cite{Fernando:2012ue}. Hussain and Ali examined the circular orbits of massive particles and determined the ISCO and OSCO for particular values of the EoS parameter, including $\omega=-1$ and $\omega=-2/3$~\cite{Hussain:2016tth}. Anand et al. investigated the stability of extremal and near-extremal Reissner–Nordström–AdS BHs in Kiselev spacetime within the framework of the Weak Gravity and Weak Cosmic Censorship Conjectures, demonstrating that the WCCC remains valid under the considered conditions~\cite{Anand:2024wxa}. Oteev et al. studied the circular motion and collisions of spinning test particles around Kerr–Kiselev BHs using the Mathisson–Papapetrou–Dixon formalism, examining the effects of quintessential dark energy on the ISCO, OSCO, and center-of-mass energy of particle collisions~\cite{Oteev:2024ewf}. Lungu et al. recently introduced an electrically charged generalization of the Kiselev BH and investigated the dynamics of charged and uncharged test particles, with particular emphasis on circular orbits, their stability, and periapsis precession~\cite{Lungu:2025pgk}. More recently, the null geodesic structure and thermodynamic properties of hairy Kiselev black holes have recently been explored, revealing the existence of stable photon orbits for various surrounding matter fields and distinctive thermodynamic features associated with cosmological horizons~\cite{Maithani:2026gad}. Beyond particle dynamics, orbital stability, and tests of the weak cosmic censorship conjecture, Kiselev BHs have also been extensively studied in the context of BH thermodynamics, gravitational lensing, and shadow observables~\cite{Shchigolev:2016gro,Xu:2016sew,Shahzad:2017vwi,Ovchinnikov:2021xfw,Cui:2021wxn,Morais:2021xmw,Abbas:2021whh,Rayimbaev:2022mrk,Nekouee:2023tew,Sood:2024ufi,Javed:2024xna,Kukreti:2025rzn,Malligawad:2025gxn,Salako:2026mdg,Ahmed:2026gpi}. Nevertheless, a systematic comparison of null and timelike circular geodesics over a broader range of $\omega$, including values outside the quintessence regime, has received comparatively little attention. In particular, the combined effect of the normalisation parameter $N$ and the EoS parameter $\omega$ on orbital stability and on the associated Lyapunov exponents requires further investigation.

The Lyapunov exponent measures the dynamical stability of an orbit through the rate at which nearby trajectories separate in phase space~\cite{Butera:1986gw,Wu:2006rx,Skokos:2008uc}. For circular geodesics, the sign of its square determines whether small radial perturbations grow or remain bounded: a positive squared Lyapunov exponent corresponds to exponential growth of the perturbation and hence an unstable orbit, whereas a negative value describes oscillations about a stable circular orbit~\cite{Setare:2010zd,Maldacena:2015waa}. Lyapunov exponents are also closely connected with BH perturbation theory, through the correspondence between unstable circular photon orbits and quasinormal-mode frequencies in the eikonal limit~\cite{PhysRevD.79.064016,Mondal:2020uwp,Giri:2021kgc,Giri:2022zhf}. They have consequently been applied widely to relativistic orbital dynamics and gravitational systems~\cite{DiCintio:2024vgp}. In this work we use the Lyapunov-exponent formalism to relate the stability of circular null and timelike geodesics quantitatively to the surrounding anisotropic matter distribution, which gives further insight into the role of $N$ and $\omega$ in the orbital dynamics of Kiselev BHs.

Motivated by these considerations, we present a systematic study of null and timelike circular geodesics around the Kiselev BHfor six representative values of $\omega$, each over its admissible range of $N$. We determine the locations and stability of circular orbits and examine how the surrounding anisotropic matter distribution modifies the orbital structure. We derive a general analytical expression for the Lyapunov exponent of circular timelike orbits, valid for arbitrary $\omega$, and study its dependence on the spacetime parameters. We show that, along the sequence of circular orbits, $(\lambda_t^T)^2$ is proportional to $-V_{\rm eff}''$, so that it changes sign exactly at the ISCO and OSCO and the Lyapunov criterion reproduces the boundaries of stable circular motion. For $\omega=-1$, $0$ and $1/3$ the metric coincides with the Schwarzschild--de Sitter metric, the Schwarzschild metric with mass $M+N/2$, and the Reissner--Nordstr\"om metric with $Q^2=-N$, respectively, and these cases provide consistency checks of our general results.

The paper is organized as follows. Section~\ref{sec:metric} introduces the Schwarzschild–Kiselev BH and discusses its horizon structure for different values of the EoS parameter $\omega$. Section~\ref{sec:geodesic} presents a detailed analysis of timelike and null geodesics and their corresponding effective potentials. Section~\ref{sec:PSISCOSA} is devoted to the photon sphere, ISCO, and orbital stability, including an investigation of stability using Lyapunov exponents. Finally, Section~\ref{sec:conclusion} summarizes our main findings and presents the conclusions. Throughout this manuscript, we adopt geometrized units in which $G=M=c=1$.

\section{Schwarzschild Kiselev Black Hole}\label{sec:metric}

In the presence of energy matter, the geometry of spacetime around a black hole is modified due to the influence of the surrounding scalar field. The resulting configuration retains spherical symmetry, but differs significantly from the vacuum Schwarzschild case. The spacetime metric can be written as \cite{kiselev2003quintessence}
\begin{equation}
ds^2 = -f(r)\,dt^2 + \frac{dr^2}{f(r)} + r^2(d\theta^2 + \sin^2\theta\, d\phi^2),
\end{equation}
with metric function
\begin{equation}\label{eq:lapsefn}
f(r) = 1 - \frac{2M}{r} - \frac{N}{r^{3\omega + 1}},
\end{equation}\\
known as the Kiselev BH spacetime; where $M$ represents the mass of the black hole, $N$ is a normalization constant associated with the field. $\omega$ is the equation-of-state parameter, characterized by an equation of state \cite{tsujikawa2013quintessence} 
\begin{eqnarray}
    \omega = \frac{p}{\rho} = \frac{\frac{\dot{\phi^2}}{2}-V(\phi)}{\frac{\dot{\phi^2}}{2}+V(\phi)},
\end{eqnarray}\label{eq:eos}\\
where $p$ is the pressure and $\rho$ is the energy density defined in terms of the kinetic energy ($\frac{\dot{\phi^2}}{2}$) and the potential energy ($V(\phi)$). Here, over dot represents the differentiation with respect to the cosmic time. If the kinetic energy term dominates, $\omega \approx 1$, if the potential energy term dominates, $\omega \approx -1$, and $\omega \approx 0$ for the same contribution of potential and kinetic energy.
The normalization constant $N$ related to the energy density as \cite{younas2015strong}
\begin{eqnarray}
    \rho = -\frac{3}{2} \frac{\omega N}{r^{3(\omega +1)}}.
\end{eqnarray}\label{eq:energyden}

\begin{figure}[h]
    \centering
    {\includegraphics[width=\linewidth]{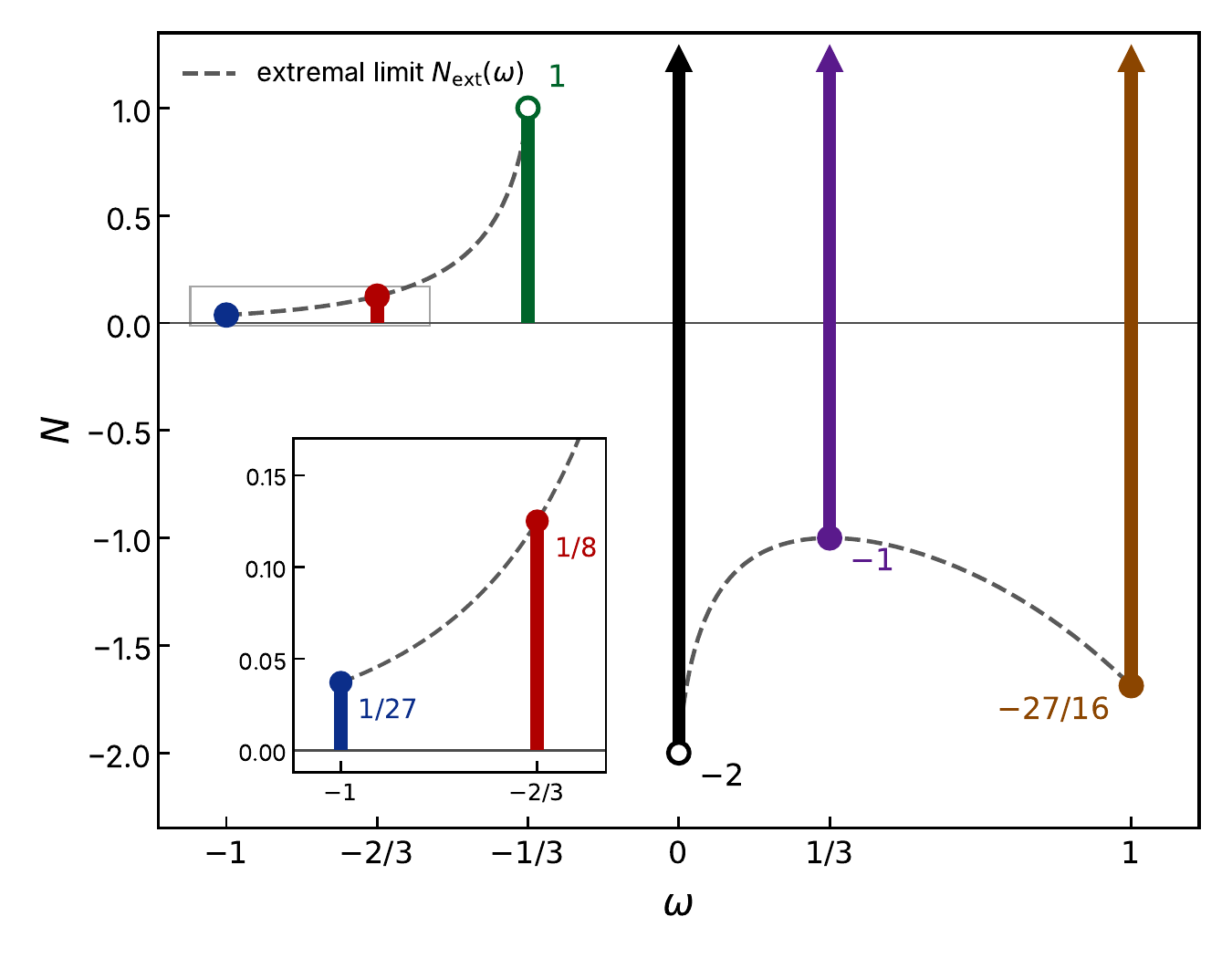}}
    \caption{Admissible range of the normalisation parameter $N$ for each value of $\omega$ considered (bars), with $M=1$. Filled and open circles denote included and excluded endpoints, respectively, and arrows indicate ranges unbounded from above. The dashed curve shows the extremal limit $N_{\rm ext}(\omega)$, at which two horizons merge; no extremal configuration exists for $-1/3<\omega<0$. The inset enlarges the region around $\omega=-1$ and $-2/3$.}
    \label{fig:NvsW}
\end{figure}

As $\rho > 0$, $\omega$ or $N$ must be negative.
If $\omega$ represents quintessence, which is a dynamical scalar field model of dark energy, it satisfies $-1 < \omega < -\frac{1}{3}$, which leads to the acceleration of the universe \cite{kiselev2003quintessence}. 
The other possible values of the state parameter may be $\omega=0$ that corresponds to pressureless matter (dust) \cite{garg2024cosmological}, $\omega=1/3$ represents radiation \cite{bouhmadi2026dark}, and $\omega=1$ represents the stiff fluid \cite{chavanis2015cosmology} as energy matter around the black hole. Different metric functions for different values of $\omega$ and the corresponding range of normalization constant $N$ are summarized in Table \ref{table:N}, and the limit of $N$ for different values of $\omega$ is shown in fig.\ref{fig:NvsW}.

The energy-momentum tensor for a Kiselev model 
\begin{equation}
T^t_{\ t} = T^r_{\ r} = \rho, \quad
T^\theta_{ \theta} = T^\phi_{\ \phi} = -\frac{1}{2}(3\omega + 1)\rho.
\end{equation}
The anisotropic form reflects the negative pressure associated with the quintessence field.

The location of the event horizon is determined by evaluating the horizon radius at which $g_{rr}=0$ or $f(r_h)$ becomes zero. The condition for the horizon radius $r_h$ gives 
\begin{equation}
    r_h^{3\omega+1}-2Mr_h^{3\omega}-N =0
\end{equation} 

\begin{figure}[htbp]
 	\centering
 	\begin{subfigure}[]{\includegraphics[width=\linewidth]{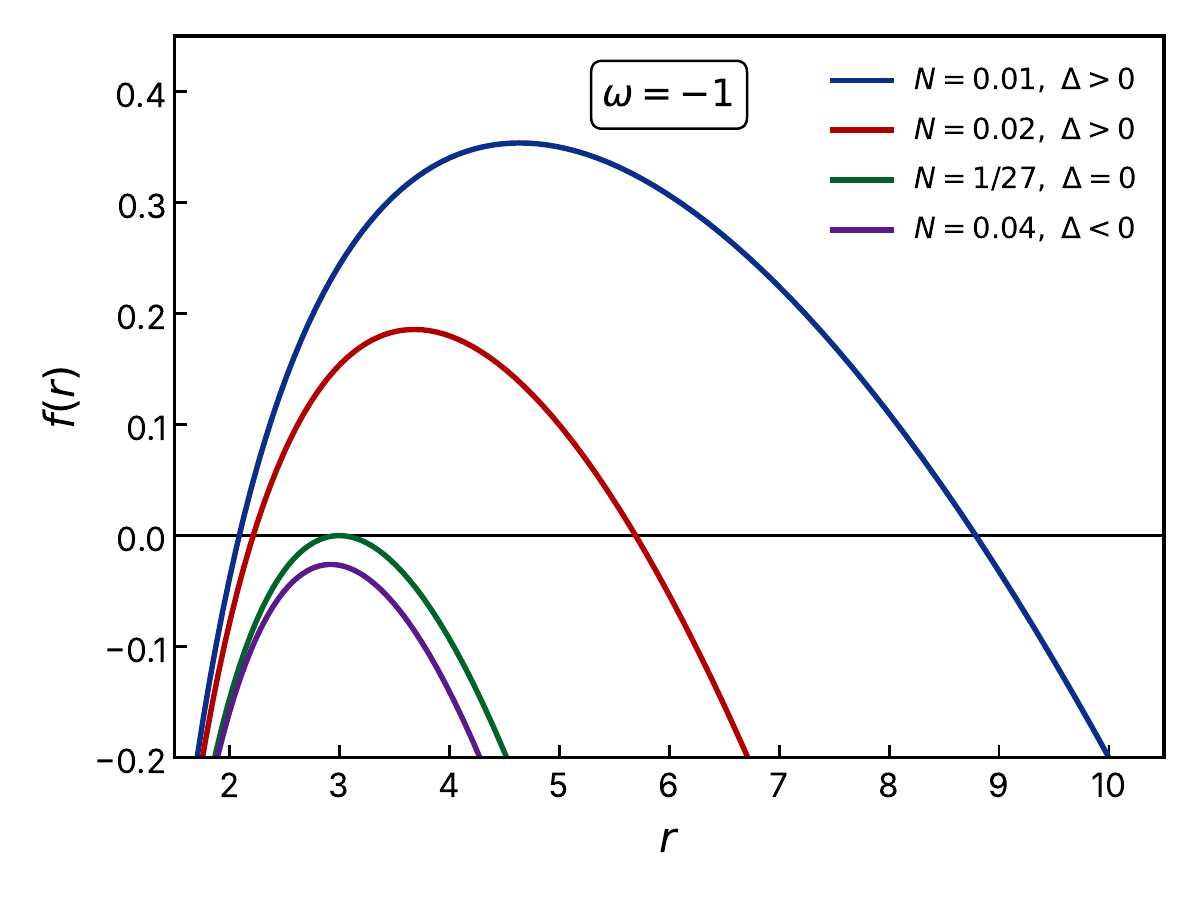}\label{fig:horizon1}}
 	\end{subfigure}
    \begin{subfigure}[]{\includegraphics[width=\linewidth]{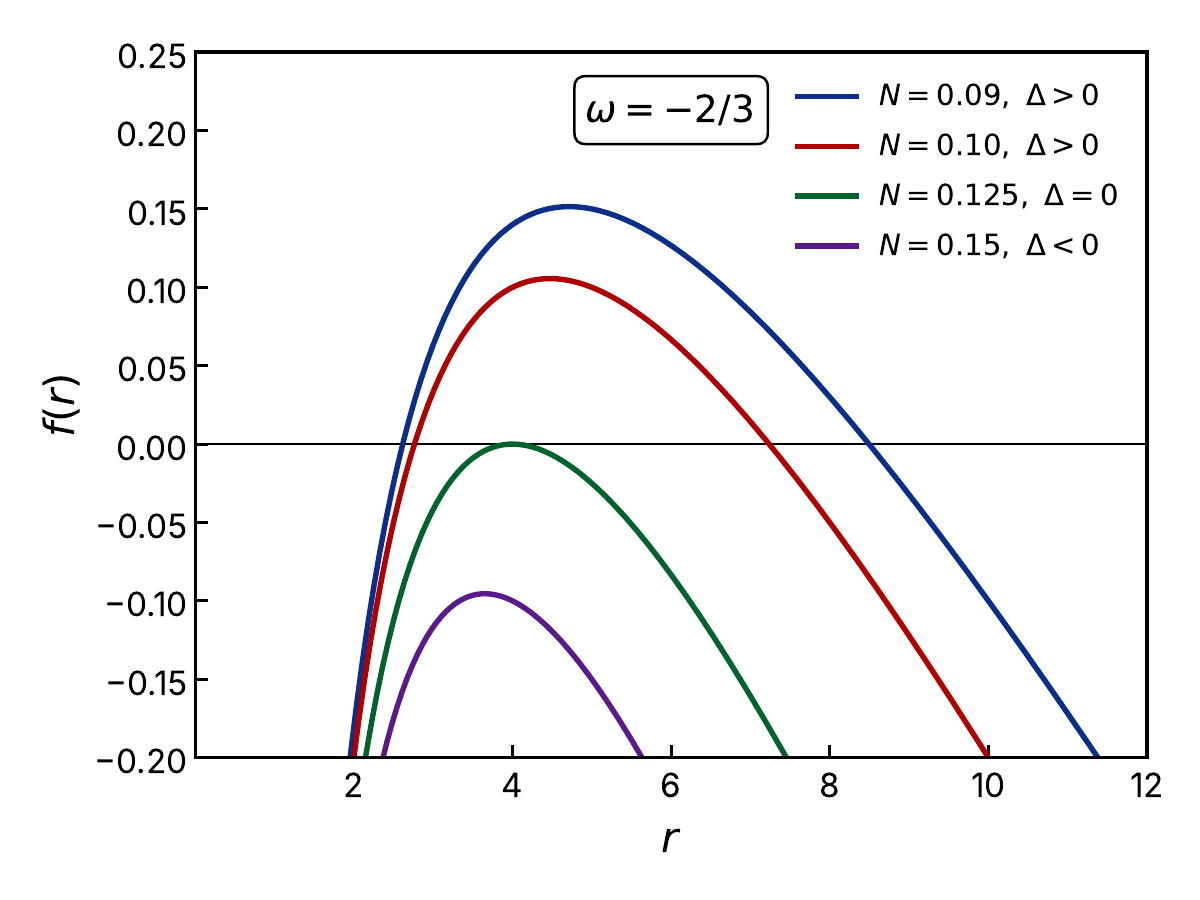}\label{fig:horizon2}}
    \end{subfigure}
    \begin{subfigure}[]{\includegraphics[width=\linewidth]{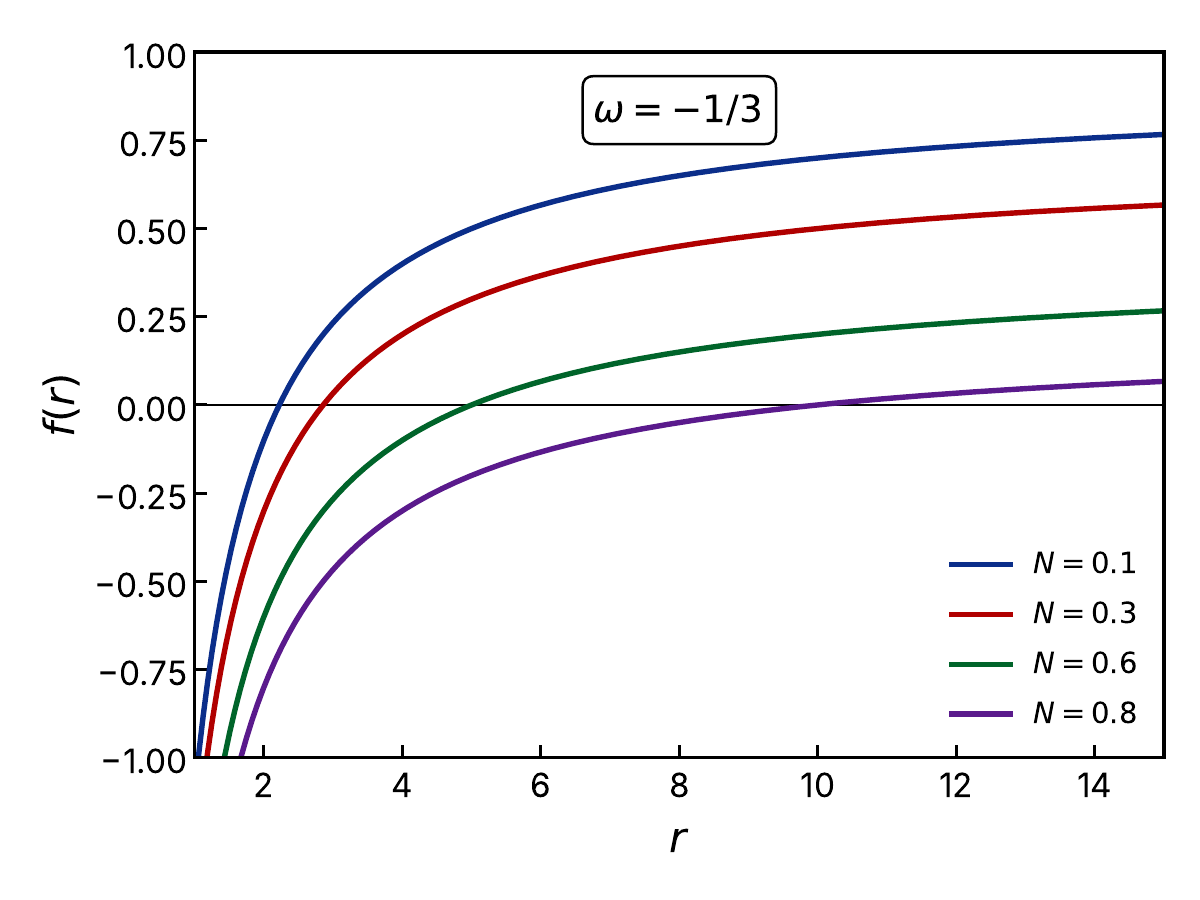}\label{fig:horizon3}}
    \end{subfigure}
 	\caption {Horizon structure variation for different values of state parameter $\omega =-1, -2/3, -1/3$ with different values of $N$ within the limiting range. }\label{fig:horizonA}
 \end{figure}

\begin{figure}[htbp]
 	\centering
 	\subfigure[]{\includegraphics[width=\linewidth]{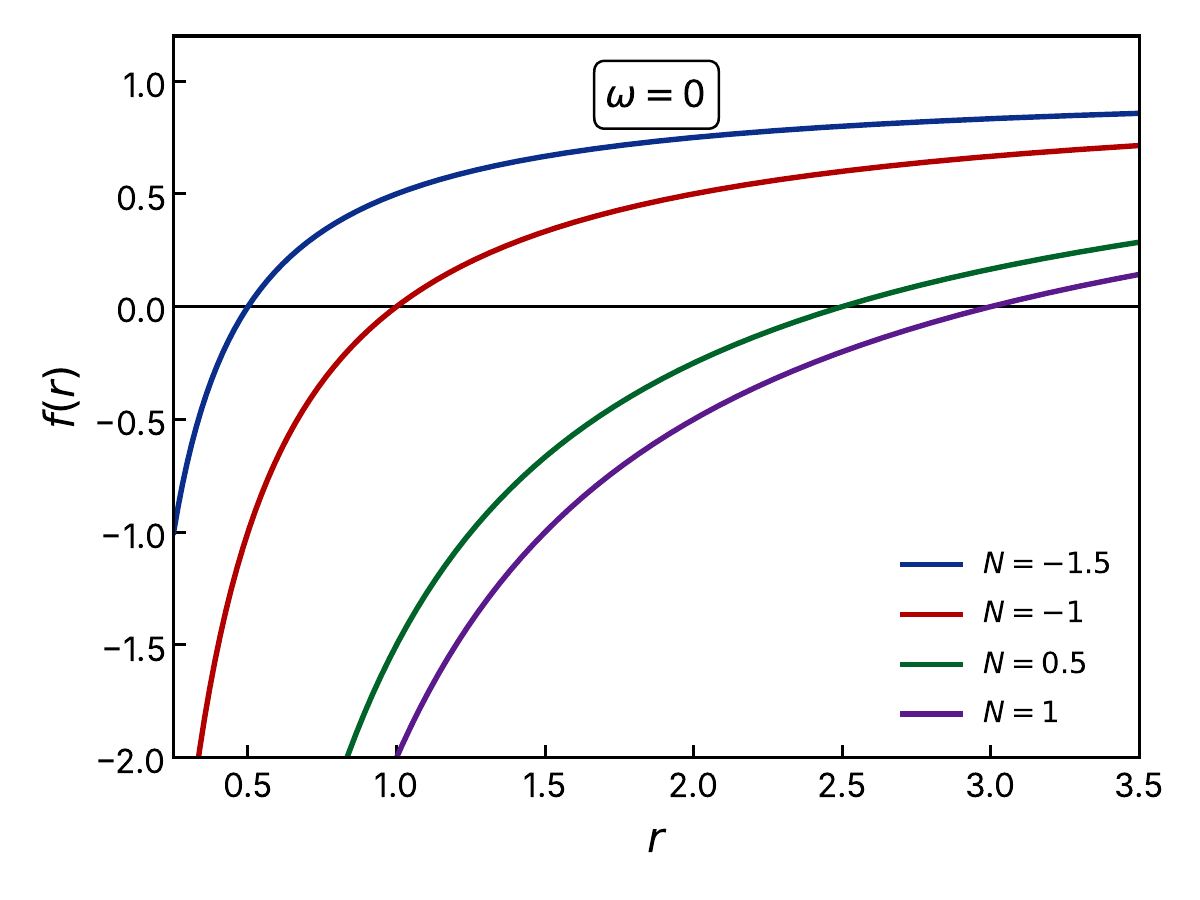}\label{fig:horizon4}}
 	\subfigure[]{\includegraphics[width=\linewidth]{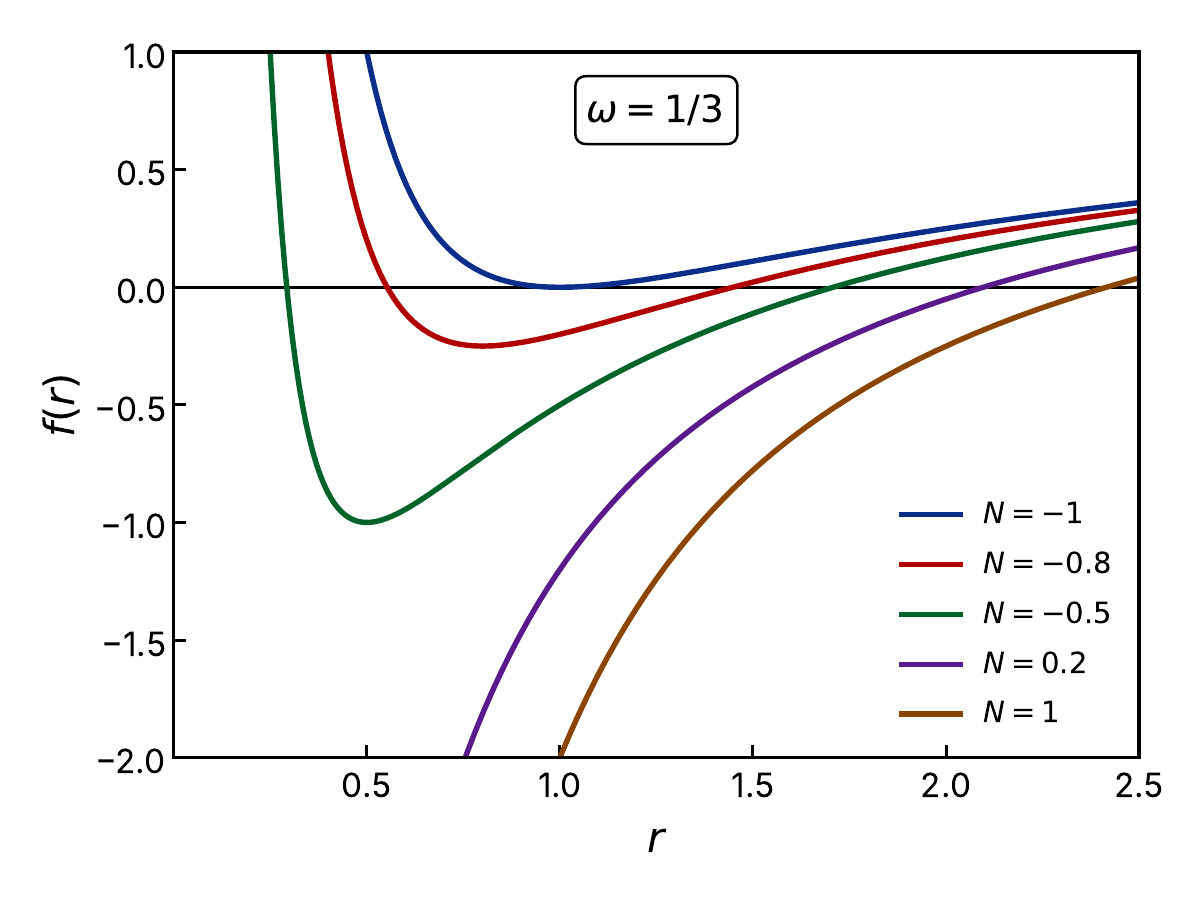}\label{fig:horizon5}}
 	\subfigure[]{\includegraphics[width=\linewidth]{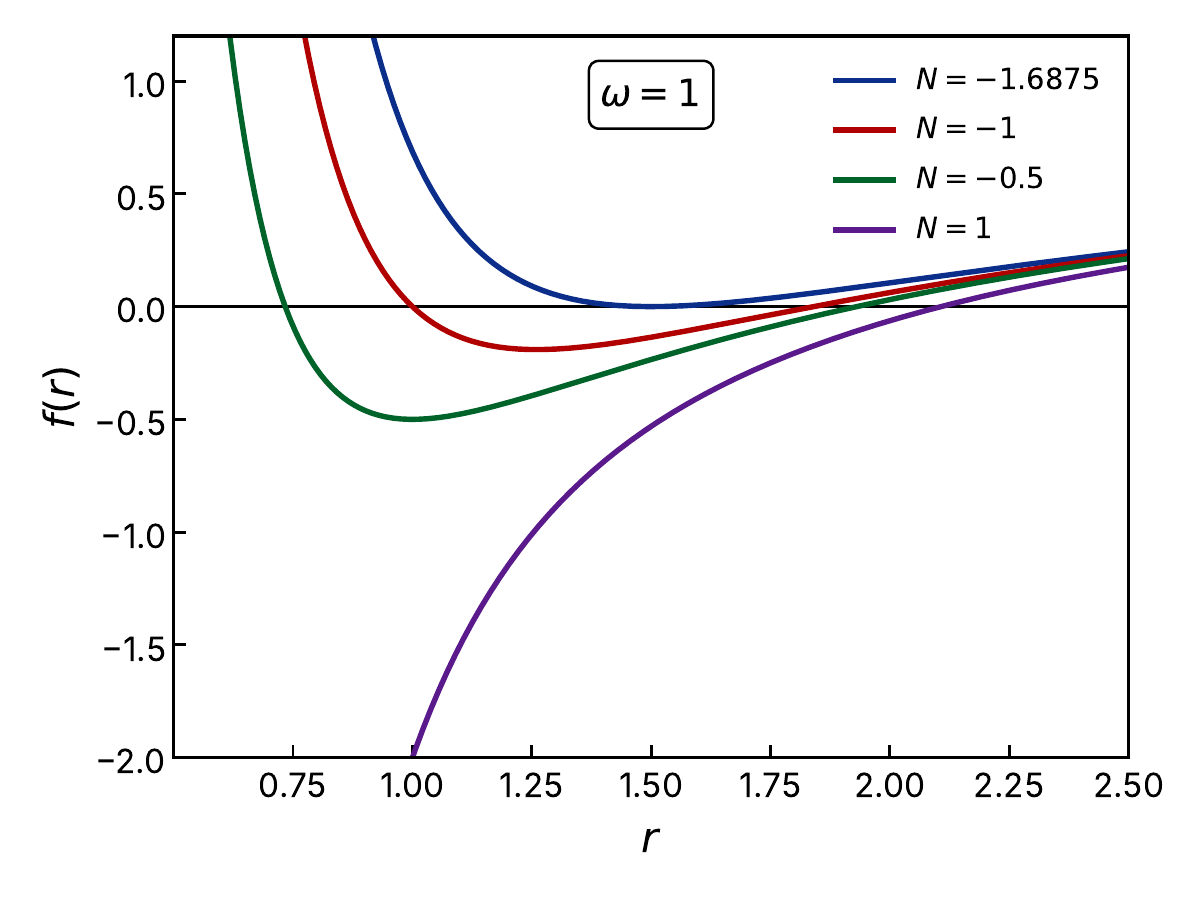}\label{fig:horizon6}}
 	\caption {Horizon structure variation for different values of state parameter $\omega = 0, 1/3, 1$ with different values of $N$ within the limiting range. }\label{fig:horizonB}
 \end{figure}

\begin{table*}[htbp]
\centering
\caption{Limit of $N$ for different values of $\omega$}
    \begin{tabular}{|c|c|c|c|c|}
   \hline
   $\omega$  & Physical interpretation & $f(r)$ &$N$ & $N ($for $M=1)$\\
     \hline
          & & & &\\
     $-1$ & Cosmological constant & $1-\frac{2M}{r}-Nr^2$ & $0 < N \leq \frac{1}{27M^2}$ & $0 < N \leq 0.037$\\
          & (de sitter type) & & &\\
     \hline
          & & & &\\
     $-2/3$ & Quintessence & $1-\frac{2M}{r}-Nr$ & $0\leq N\leq\frac{1}{8M}$ & $0\leq N\leq0.125$\\
     & & &&\\
     \hline
          & & & &\\
     $-1/3$ &Borderline Quintessence & $1-\frac{2M}{r}-N$ & $0 \leq N < 1$& $0 \leq N < 1$\\
          & & &&\\
     \hline
          & & & &\\
     $0$ & Pressureless matter  & $1-\frac{2M}{r}- \frac{N}{r}$ & $-2<N $& $-2<N $\\
          &  (Dust) & & &\\
     \hline
          & & & &\\
     $\frac{1}{3}$ & Radiation & $1-\frac{2M}{r}-\frac{N}{r^2}$ & $-1\leq N$ & $-1\leq N$\\
          & & & &\\
     \hline
          & & & &\\
     $1$ & Stiff fluid &$1-\frac{2M}{r}-\frac{N}{r^4}$ & $-\frac{27M^4}{16}\leq N$& $-1.6875 \leq N$\\
          & & & &\\
     \hline
     \end{tabular} \label{table:N}
\end{table*}
The nature of the horizons for different values of $N$ and $\omega$ is shown in figs. \ref{fig:horizonA} $\&$ \ref{fig:horizonB} and discussed below:
\begin{itemize}
    \item With $\omega=0$ and $N=0$ we recover the Schwarzschild BH solution $f(r)=1 - \frac{2M}{r}$ without any external field, which corresponds to a single horizon $r_h=2M$.
    \item $\omega=-1$ corresponds to the Schwarzschild de-Sitter type solution $f(r)=1 - \frac{2M}{r} - N r^2$, has the limit $N \leq 0.037$ for $M=1$, gives one negative root (non-physical) and two positive real roots (physical horizons), shown in fig. \ref{fig:horizon1}. For the discriminant $\Delta > 0$, the value of $N$ is 0.01 and there exist two horizons, the BH horizon and the cosmological horizon. 
    When $\Delta$ becomes zero, this is the condition of the extremal horizon and in the case of $\Delta < 0$, there is no existence of the physical horizon.
    \item $\omega=-2/3$ gives the metric function $f(r)=1 - \frac{2M}{r} - N r $ as the Schwarzschild metric surrounded by quintessence matter has horizon at
\begin{equation*}
    r_{h}^{\pm} =\frac{1\pm\sqrt{1-8NM}}{2N},
\end{equation*}
    has the limit $N \leq 0.125$ for $M=1$, gives two horizons as event horizon and quintessence horizons, shown in fig. \ref{fig:horizon2}. From this horizon plot, it is evident that for $N=0.1$,  $\Delta > 0$ that gives two horizons. $r_{h}^{-}$ implies an event horizon, and $r_{h}^{+}$ corresponds to the quintessence horizon. For $N = 0.125$, $\Delta=0$ implies the extremal horizon and beyond the limit of $N$, there is no existence of the physical horizon.  
    \item $\omega=-1/3$  represents the boundary of the Quintessence in the Schwarzschild metric gives the lapse function as $f(r)=1-\frac{2M}{r}-N$, has the limit of $N < 1$ for $M = 1$ and gives a horizon at
\begin{equation*}
    r_h= \frac{2M}{1-N}.
\end{equation*}
In fig. \ref{fig:horizon3}, we see that as $N$ increases, the zeros of $f(r)$ shift towards the larger value of $r$, which implies the strong influence of quintessence on spacetime structure.
    \item With $\omega=0$ but $N \neq 0$ we recover the metric function as $f(r)=1-\frac{2M}{r}- \frac{N}{r}$ has horizon at
\begin{equation*}
    r_h=2M+N,    
\end{equation*}
but no horizon in the limiting value of $N=-2$, for $M=1$ which is explicitly shown in fig. \ref{fig:horizon4}. From this figure it is also evident that only one horizon exists for $N > -2$ but for $N=-1, 0.5, 1.0$,  $f(r)$ gradually approaches zero at a higher value of $r$, which indicates that spacetime is evolving from the strong field region towards the outer region where the influence of $N$ is weak. And the parameter $N$ has a significant impact on the radial geometry and the horizon structure.
    \item When $\omega=1/3$, the metric function becomes $f(r)=1-\frac{2M}{r}-\frac{N}{r^2}$, has horizon at
\begin{equation*}
    r_h^{\pm}= M \pm \sqrt{M^2+N},
\end{equation*}
where  $N = -1$ is the limit value for having the extremal horizon located at $r_h=1$. From fig. \ref{fig:horizon5} we observe that for $N=-0.5$, there exist two horizons but for $N=0.5, 1.0$, only one horizon exists. From the horizon expression, it is evident that for $-1\leq$ $N<0$, there are two horizons and for positive value of $N$ ($N>0$) only one horizon is possible. Also, at smaller $r$ the influence of $N$ (specifically for negative values) is very strong, whereas at larger $r$ the influence of $N$ becomes weak so that no such huge variation occurs in the outer region.
   \item With $\omega=1$, $f(r)=1-\frac{2M}{r}-\frac{N}{r^4}$. Fig. \ref{fig:horizon6} shows that $N = -1.6875$ is the limiting range of $N$ for having an extremal horizon. Unlike $\omega=1/3$, although the range of $N$ is allowed for positive values, there is only one horizon for $N > 0 $ and two horizons for $-1.6875 \leq N < 0$.
\end{itemize}


\section{Geodesics and effective potentials}\label{sec:geodesic}
The motion of a test particle in a curved spacetime can be derived from the variational principle. The Lagrangian  for the motion is
\begin{equation}\label{eq:lagrangian}
\mathcal{L} = \frac{1}{2} g_{\mu\nu} \dot{x}^\mu \dot{x}^\nu,
\end{equation}
where an over dot ($^.$) implies differentiation with respect to the affine parameter $\tau$.
Elaborating Eq. \ref{eq:lagrangian} by inserting Eq. \ref{eq:lapsefn} we get the corresponding Lagrangian in the equatorial plane ($\theta=\pi/2$) as
\begin{equation}
    \mathcal{L}= \frac{1}{2}\left[-f(r)\dot{t}^{2}+\frac{\dot{r}{^2}}{f(r)}+r^2\sin^2\theta\dot{\phi}{^2}\right].
\end{equation}
Because of the time translational symmetry and rotational symmetry of the spacetime, the corresponding momenta are constants of motion, expressed as
\begin{equation}
    \frac{\partial\mathcal{L}}{\partial\dot{t}}=-f(r)\dot{t}=-E,  \quad \frac{\partial\mathcal{L}}{\partial\dot{\phi}}=r^2\dot{\phi}=L, \quad
     \frac{\partial\mathcal{L}}{\partial\dot{r}}=\frac{\dot{r}}{f(r)};
\end{equation}
which yields
\begin{equation}
    p_t=-f(r)\dot{t}=-E, \quad p_{\phi}=r^2\dot{\phi}=L, \quad p_r=\frac{\dot{r}}{f(r)};
\end{equation}
or, 
\begin{equation}\label{eq:consrvd}
    \dot{t}=\frac{E}{f(r)}, \quad \dot{\phi}=\frac{L}{r^2}, \quad \dot{r}=p_r f(r) .
\end{equation}
With the concept of the Euler Lagrangian equation, geodesic equations for time coordinate ($t$), radial coordinate ($r$), and azimuthal coordinates ($\phi$) are as follows:
\begin{equation}
    \ddot{t}+\frac{f'(r)}{f(r)}\dot{r}\dot{t}=0,
\end{equation}
\begin{equation}
    \ddot{r}+\frac{f'(r)f(r)}{2}\dot{t^2}-\frac{f'(r)}{2f(r)}\dot{r^2}-rf(r)\dot{\phi^2}=0,
\end{equation}
\begin{equation}
    \ddot{\phi}+\frac{2}{r}\dot{r}\dot{\phi}=0.
\end{equation}
The geodesics of a test particle are constraint by the condition

\begin{equation}\label{eq:constraint}
    g_{\mu\nu} \dot{x}^\mu \dot{x}^\nu=-\kappa,
\end{equation}

where $\kappa= 1$ implies the timelike geodesic and $\kappa=0$ for null geodesics in the metric signature ($-,+,+,+$). 
Substituting the conserved quantities from Eq. (\ref{eq:consrvd}) into the constraint Eq. (\ref{eq:constraint}) the radial equation can be written in the form 
\begin{equation}\label{}
     \dot{r^2}=E^2-f(r)\left(\kappa+\frac{L^2}{r^2}\right).
 \end{equation}
Compared with the Newtonian central force problem, the above equation also reads as
\begin{equation}
    \dot{r^2}=E^2 - V_{eff}(r),
\end{equation}
and
\begin{equation}\label{eq:ddotr}
    \ddot{r}= - \frac{1}{2}V_{eff}'(r)
\end{equation}
where the effective potential is represented as 
\begin{eqnarray}\label{eq:effective}
    V_{eff}(r)&=&f(r)\left(\kappa+\frac{L^2}{r^2}\right)\\
   & = &\left(1 - \frac{2M}{r} - \frac{N}{r^{3\omega + 1}} \right) \left(\kappa+\frac{L^2}{r^2}\right).
\end{eqnarray}
From the above expression, it is evident that the effective potential depends on the BH mass ($M$), the normalization constant ($N$), and the state parameter ($\omega$). However, the angular momentum ($L$) also changes the effective potential because it produces a stronger centrifugal barrier and affects the radial motion of the particle.
\subsection{Analysis of effective potential for timelike geodesics with different values of \texorpdfstring{$\omega$}{omega}}
\begin{figure*}[htbp]
	\centering
	\subfigure[]{\includegraphics[scale=0.33]{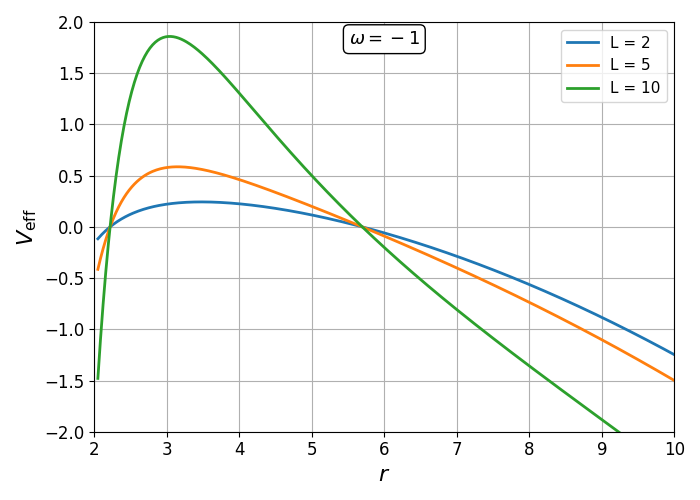}\label{fig:efT1}}
    \subfigure[]{\includegraphics[scale=0.33]{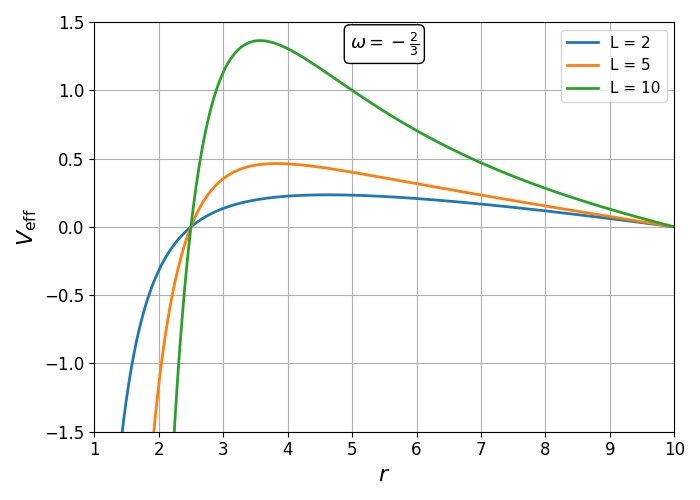}\label{fig:efT3}}
    \subfigure[]{\includegraphics[scale=0.33]{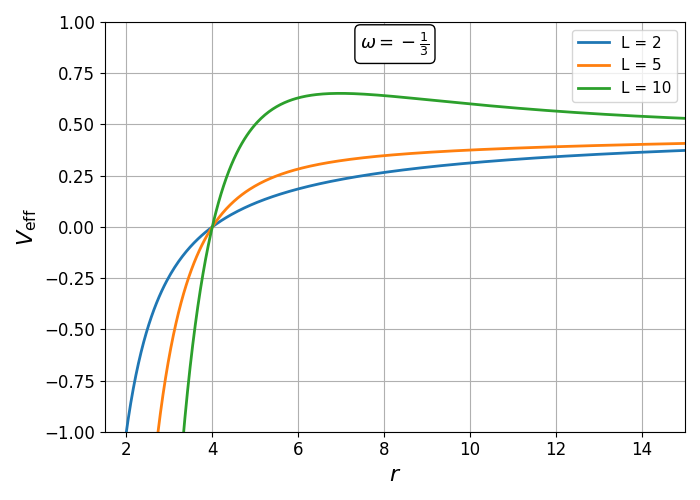}\label{fig:efT5}}
   	\subfigure[]{\includegraphics[scale=0.33]{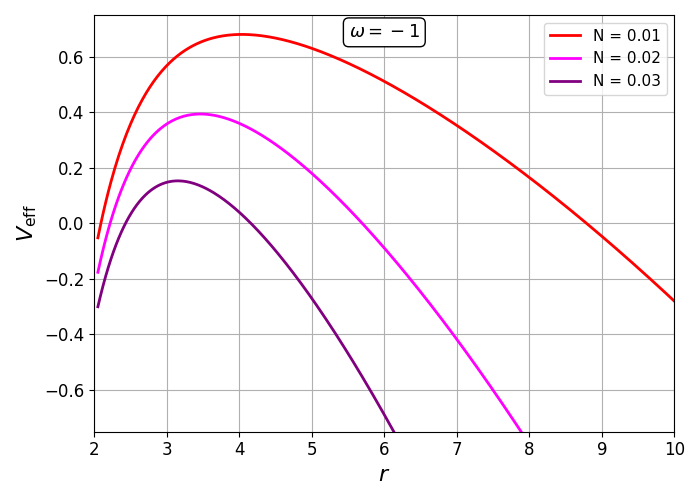}\label{fig:efT2}}
    \subfigure[]{\includegraphics[scale=0.33]{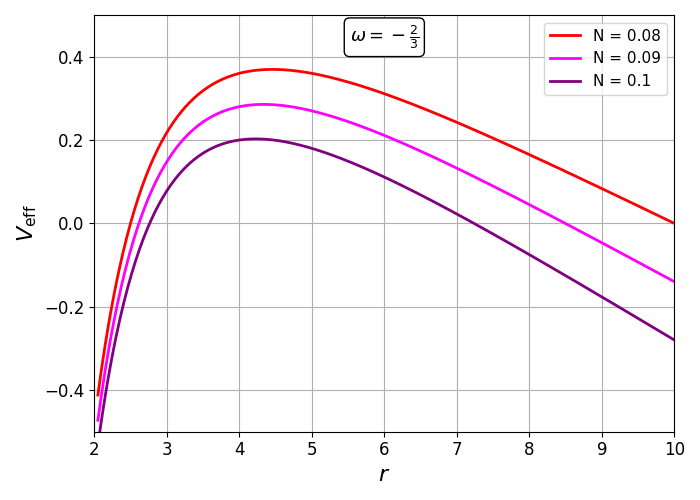}\label{fig:efT4}}
    \subfigure[]{\includegraphics[scale=0.33]{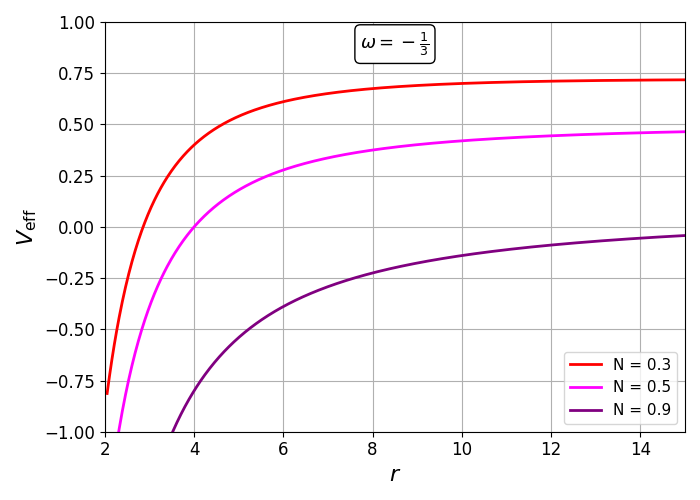}\label{fig:efT6}}
	\caption {Variation of effective potential $v_{eff}$ of timelike geodesics with $r$ for $\omega=-1, -2/3, -1/3$. In the upper panel, the value of $N$ kept constant as $N=0.02, 0.08, 0.5$ respectively while the angular momentum $L$ varies as $L=2,5,10$. In the lower panel, the value of $L$ remains constant as $L=4$ while $N$ varies depending on the value of $\omega$.}\label{fig:EPT1}
\end{figure*}

\begin{figure*}[htbp]
	\centering
	\subfigure[]{\includegraphics[scale=0.33]{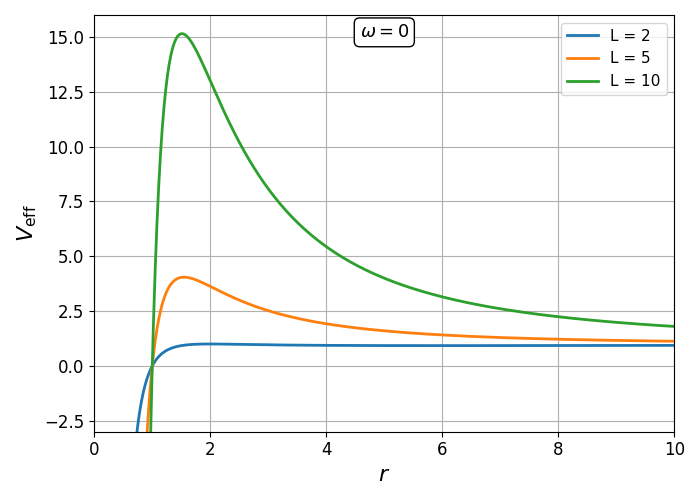}\label{fig:efT7}}
    \subfigure[]{\includegraphics[scale=0.33]{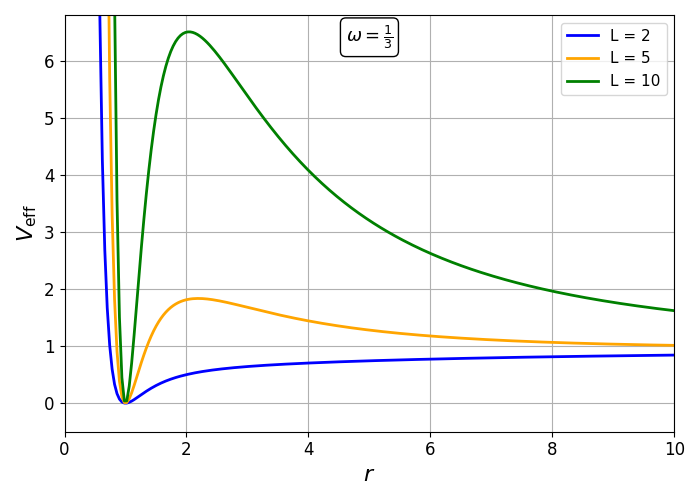}\label{fig:efT9}}
    \subfigure[]{\includegraphics[scale=0.33]{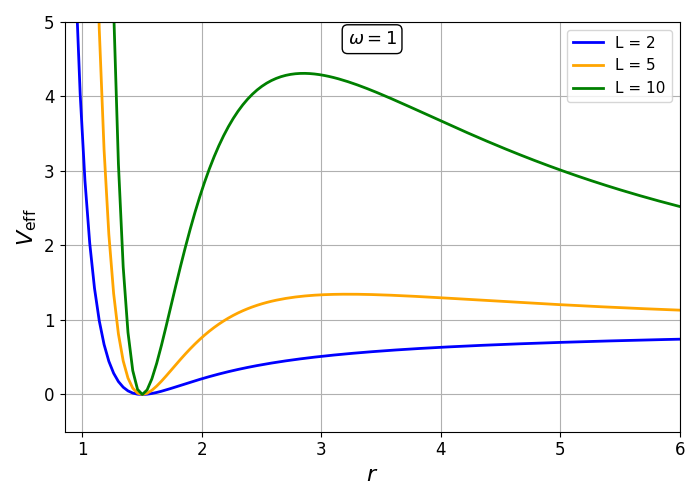}\label{fig:efT11}}
   	\subfigure[]{\includegraphics[scale=0.33]{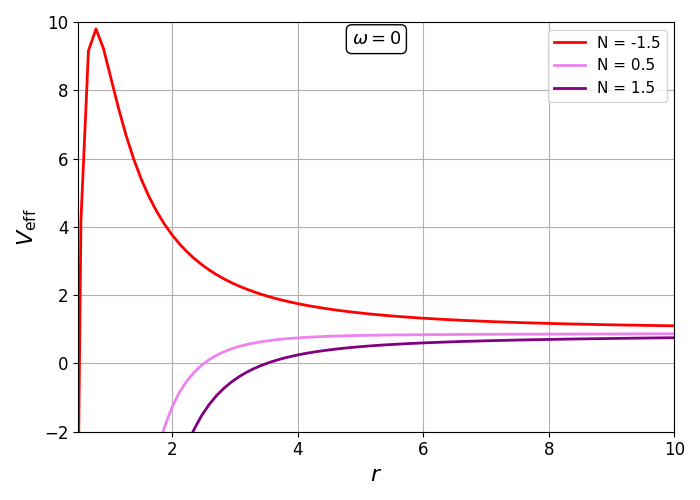}\label{fig:efT8}}
    \subfigure[]{\includegraphics[scale=0.33]{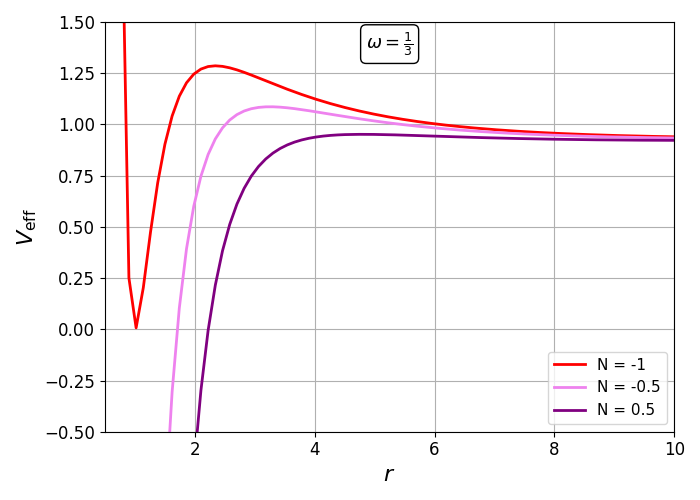}\label{fig:efT10}}
    \subfigure[]{\includegraphics[scale=0.33]{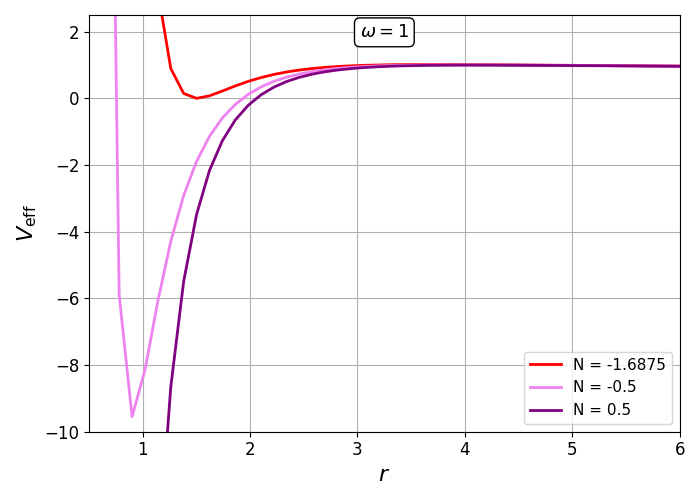}\label{fig:efT12}}
	\caption {Variation of effective potential $v_{eff}$ of timelike geodesics with $r$ for $\omega=0, 1/3, 1$. In the upper panel, the value of $N$ kept constant as $N= -1, -1, -1.6875$ respectively while the angular momentum $L$ varies as $L=2,5,10$. In the lower panel, the value of $L$ remains constant as $L=4$ while $N$ varies depending on the value of $\omega$.}\label{fig:EPT2}
\end{figure*}

The effective potential for timelike geodesics can be obtained by considering ($\kappa=1$) as
\begin{equation}\label{eq:effective_time}
    V_{eff}^T(r)=\left(1 - \frac{2M}{r} - \frac{N}{r^{3\omega + 1}} \right) \left(1+\frac{L^2}{r^2}\right).
\end{equation}
Now, with different values of the state parameter $\omega$, the effective potential can be obtained for timelike geodesics.
Figs.~\ref{fig:EPT1} and \ref{fig:EPT2} show the effective potential for timelike geodesics as a function of the radial coordinate $r$ for $\omega=-1$, $-2/3$, $-1/3$ and $\omega=0$, $1/3$, $1$, respectively. For each $\omega$, the effective potential is plotted for several values of $L$ and $N$, with $N$ chosen within the admissible ranges given in Table~\ref{table:N}. For fixed $N$, increasing $L$ raises the peak of the effective potential and moves it inwards, towards the photon sphere, for every value of $\omega$. For fixed $L$, increasing $N$ lowers the effective potential at every radius. However, the effect of $N$ on the location of the peak depends on $\omega$. The peak moves inwards for $\omega=-1$ and outwards for $\omega=-1/3$, $0$, $1/3$ and $1$, while for $\omega=-2/3$ it moves inwards for $L<4$, remains at $r=4$ for $L=4$, and moves outwards for $L>4$. For small $L$, increasing $N$ can also completely remove the extrema, so there is no circular orbit for that angular momentum. 

Table~\ref{tab:isco_veff} shows that, for all combinations of $L$ and $N$ considered, the effective potential for $\omega=-1$ and $-2/3$ has a single maximum and no minimum, so that no stable circular orbits exist for these parameter values. This is because all the values of $N$ used here exceed the critical value $N_c$ obtained in Sec.~\ref{sec:PSISCOSA}; for $N<N_c$, stable circular orbits do exist between the ISCO and the OSCO. In contrast, for $\omega=-1/3$, $0$, $1/3$ and $1$, the effective potential develops one maximum and one minimum for sufficiently large $L$, corresponding to an unstable and a stable circular orbit, respectively.

\begin{figure*}[htbp]
	\centering
	\subfigure[]{\includegraphics[scale=0.33]{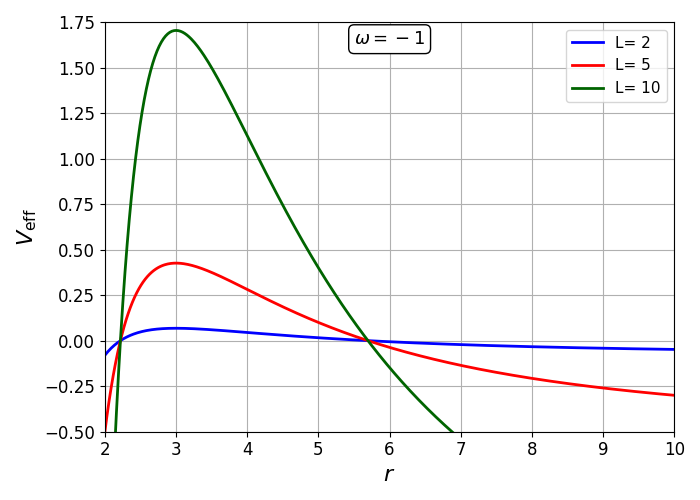}}\label{fig:efT13}
    \subfigure[]{\includegraphics[scale=0.33]{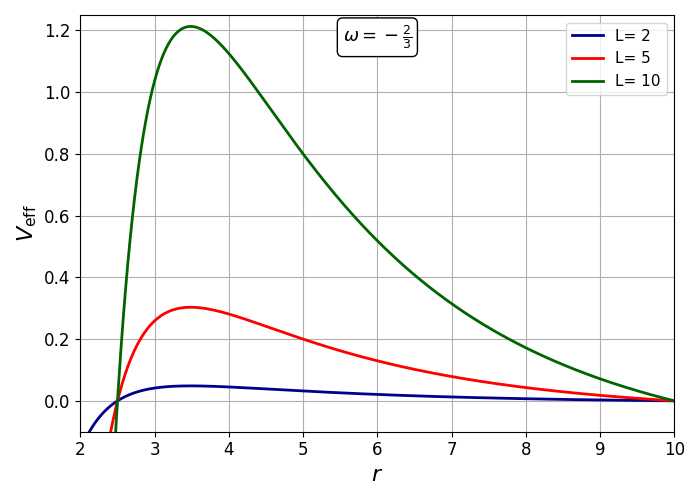}\label{fig:efT15}}
    \subfigure[]{\includegraphics[scale=0.33]{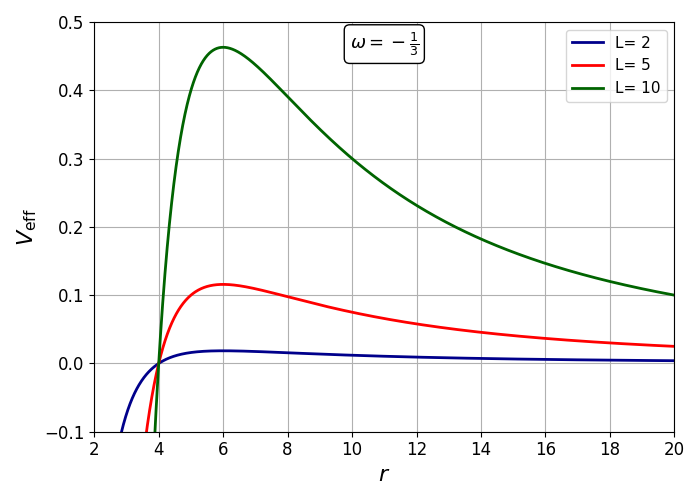}\label{fig:efT17}}
   	\subfigure[]{\includegraphics[scale=0.33]{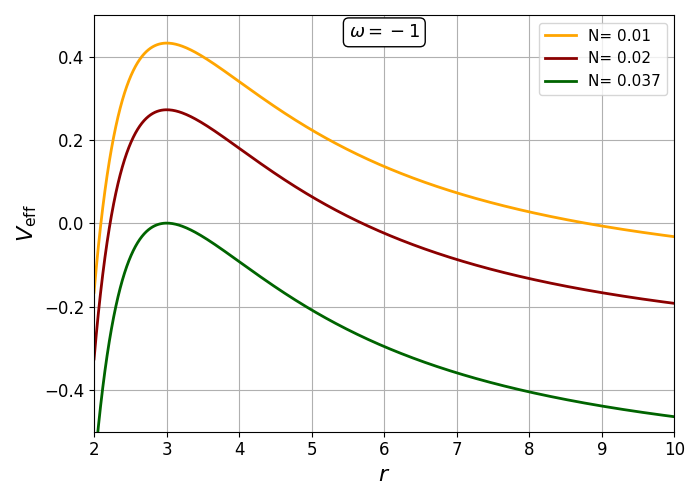}\label{fig:efT14}}
    \subfigure[]{\includegraphics[scale=0.33]{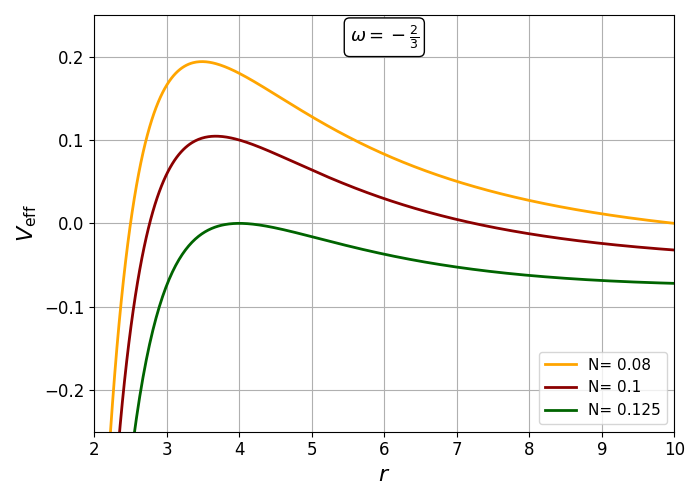}\label{fig:efT16}}
    \subfigure[]{\includegraphics[scale=0.33]{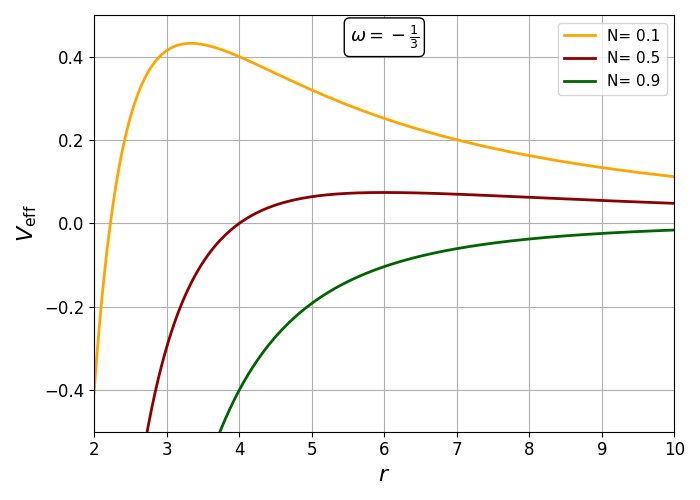}\label{fig:efT18}}
	\caption {Variation of effective potential $v_{eff}$ of null geodesics with $r$ for $\omega=-1, -2/3, -1/3$. In the upper panel, the value of $N$ kept constant as $N=0.02, 0.08, 0.5$ respectively while the angular momentum $L$ varies as $L=2,5,10$. In the lower panel, the value of $L$ remains constant as $L=4$ while $N$ varies depending on the value of $\omega$.}\label{fig:EPN1}
\end{figure*}
\begin{figure*}[htbp]
	\centering
	\subfigure[]{\includegraphics[scale=0.33]{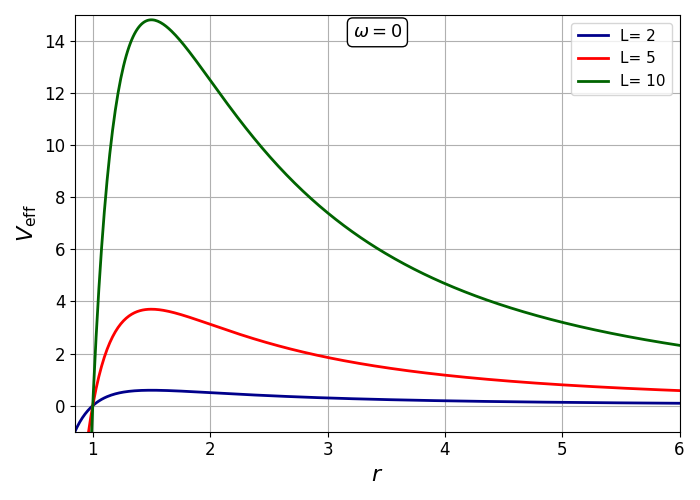}\label{fig:efT19}}
    \subfigure[]{\includegraphics[scale=0.33]{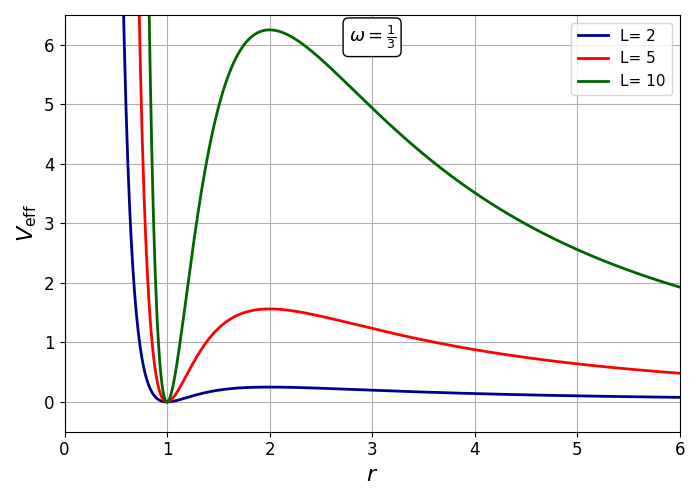}\label{fig:efT21}}
    \subfigure[]{\includegraphics[scale=0.33]{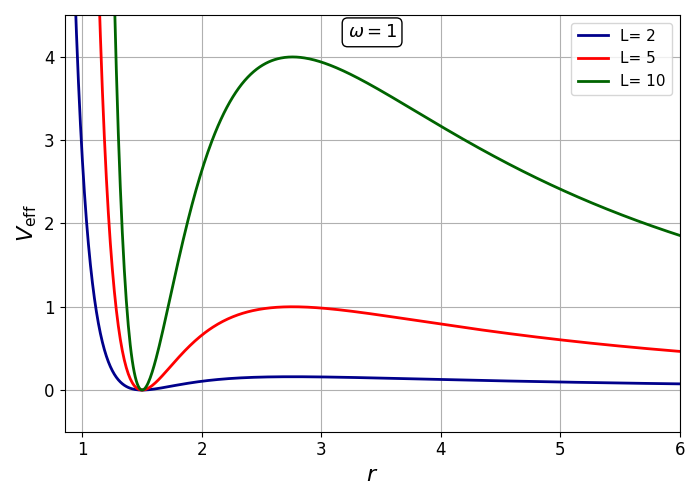}\label{fig:efT23}}
   	\subfigure[]{\includegraphics[scale=0.33]{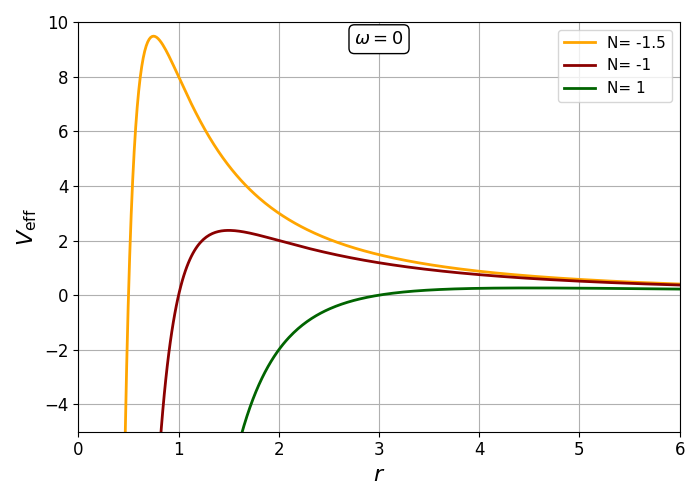}\label{fig:efT20}}
    \subfigure[]{\includegraphics[scale=0.33]{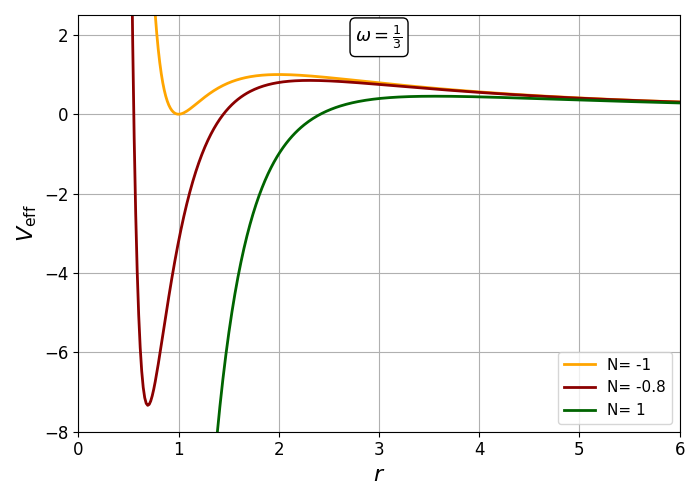}\label{fig:efT22}}
    \subfigure[]{\includegraphics[scale=0.33]{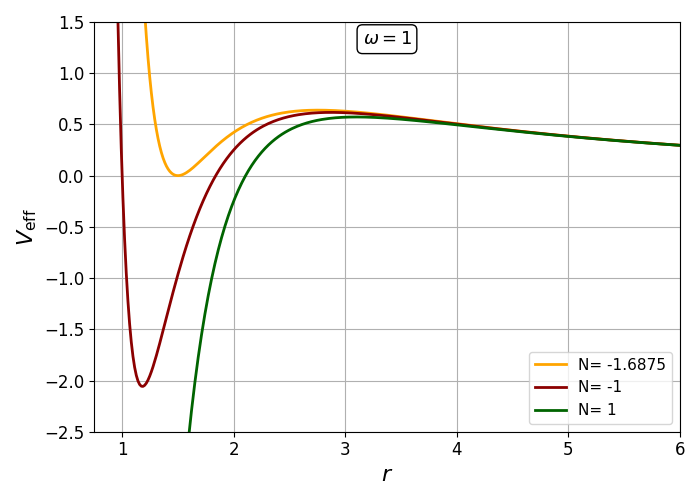}\label{fig:efT24}}
	\caption {Variation of effective potential $v_{eff}$ of null geodesics with $r$ for $\omega=0, 1/3, 1$. In the upper panel, the value of $N$ kept constant as $N= -1, -1, -1.6875$ respectively while the angular momentum $L$ varies as $L=2,5,10$. In the lower panel, the value of $L$ remains constant as $L=4$ while $N$ varies depending on the value of $\omega$.}\label{fig:EPN2}
\end{figure*}

\begin{table*}[htbp]
\centering
\setlength{\tabcolsep}{4pt}
\renewcommand{\arraystretch}{1.25}
\caption{ISCO radius and the radial positions corresponding to the maximum and minimum of the effective potential $V_{\rm eff}$ for different values of $\omega$, $N$, and $L$, with $M=1$. Here, $r(V_{\rm eff}^{\max})$ and $r(V_{\rm eff}^{\min})$ denote the values of $r$ at which the effective potential reaches its maximum and minimum, respectively. Only physically admissible extrema in the static region are retained.}
\label{tab:isco_veff}
\begin{tabular}{|c|c|c|cccc|cccc|}
\hline
\multirow{2}{*}{$\omega$} &
\multirow{2}{*}{$N$} &
\multirow{2}{*}{$r_{\rm ISCO}$} &
\multicolumn{4}{c|}{$r$ at $V_{\rm eff}^{\max}$} &
\multicolumn{4}{c|}{$r$ at $V_{\rm eff}^{\min}$}
\\
& 
& 
& $L=2$ & $L=4$ & $L=5$ & $L=10$
& $L=2$ & $L=4$ & $L=5$ & $L=10$
\\
\hline
\multirow{3}{*} {-1} & $0.01$
& --
& 4.17922 & 3.43854 & 3.27843 & 3.06693
& -- & -- & -- & --
\\
& $0.02$
& --
& 3.47852 & 3.21631 & 3.14894 & 3.04048
& -- & -- & -- & --
\\
& $0.03$
& --
& 3.15147 & 3.07533 & 3.05421 & 3.01609
& -- & -- & -- & --
\\
\hline
\multirow{3}{*} {-2/3} & $0.01$
& --
& 12.46008 & 4.00000 & 3.52956 & 3.14333
& -- & -- & -- & --
\\
& $0.08$
& --
& 4.62618 & 4.00000 & 3.82897 & 3.57317
& -- & -- & -- & --
\\
& $0.09$
& --
& 4.44149 & 4.00000 & 3.86906 & 3.65418
& -- & -- & -- & --
\\
\hline
\multirow{3}{*} {-1/3} & $0.3$
& 8.57143 
& -- & -- & 7.50000 & 4.58619
& -- & -- & 10.00000 & 65.41381
\\
& $0.5$
& 12.00000 
& -- & -- & -- & 6.97224
& -- & -- & -- & 43.02776
\\
& $0.9$
& 60.00000 
& -- & -- & -- & --
& -- & -- & -- & --
\\
\hline
\multirow{4}{*} {0} & $-1.5$
& 1.50000
& 0.78890 & 0.75900 & 0.75571 & 0.75141
& 15.21110 & 63.24100 & 99.24429 & 399.24859
\\ 
& $-1$
& 3.00000 
& 2.00000 & 1.57779 & 1.54792 & 1.51142
& 6.00000 & 30.42221 & 48.45208 & 198.48858
\\
& $0.5$
& 7.50000 
& -- & -- & 5.00000 & 3.94449
& -- & -- & 15.00000 & 76.05551
\\
& $1.5$
& 10.50000 
& -- & -- & -- & 5.84861
& -- & -- & -- & 51.29425
\\
\hline
\multirow{3}{*} {1/3} & $-1$
& 4.00000 
& -- & 2.34315 & 2.19224 & 2.04168
& 1.00000 & 13.65685 & 22.80776 & 97.95832
\\
& $-0.5$
& 5.16646 
& -- & 3.25170 & 2.95017 & 2.68651
& -- & 12.86585 & 22.16755 & 97.43145
\\
& $0.5$
& 6.69451 
& -- & 4.79596 & 3.95861 & 3.42639
& -- & 11.00713 & 20.84437 & 96.37644
\\
\hline
\multirow{3}{*} {1} & $-1.6875$
& 5.71268 
& -- & 3.69699 & 3.22200 & 2.85098
& 1.50000 & 12.04063 & 21.52328 & 96.90453
\\
& $-0.5$
& 5.92135 
& -- & 3.91908 & 3.41806 & 3.03518
& -- & 12.01212 & 21.51666 & 96.90427
\\
& $0.5$
& 6.07434 
& -- & 4.07570 & 3.54852 & 3.15040
& -- & 11.98781 & 21.51109 & 96.90405
\\
\hline
\end{tabular}\label{table:ISCO}
\end{table*}

\subsection{Analysis of effective potential for null geodesics for different values of \texorpdfstring{$\omega$}{omega}}
The effective potential for null geodesics is obtained  by setting $\kappa=0$,
\begin{equation}\label{eq:effective_null}
    V_{eff}^N(r)=\left(1 - \frac{2M}{r} - \frac{N}{r^{3\omega + 1}} \right) \left(\frac{L^2}{r^2}\right).
\end{equation}
Using Eq. (\ref{eq:effective_null}) the effective potential can be obtained for null geodesics with different values of the state parameter.
Figs.~\ref{fig:EPN1} and \ref{fig:EPN2} show $V_{\rm eff}^N$ as a function of the radial coordinate $r$ for $\omega=-1$, $-2/3$, $-1/3$ and $\omega=0$, $1/3$, $1$, respectively, for several values of $L$ and of $N$ within the admissible ranges given in Table~\ref{table:N}. The change in the value of angular momentum only rescales the potential and does not shift the location of its extrema. In the region outside the event horizon, and inside the cosmological horizon where one exists, the potential has a single maximum for every $\omega$ and $N$, located at the photon sphere, so that circular photon orbits are unstable in all cases considered. For $\omega=1/3$ and $1$ with $N<0$, the curves also show a minimum, but it lies either on the degenerate horizon in the extremal case or between the inner and outer horizons, where $f(r)<0$. It therefore does not correspond to a physical circular orbit, and no stable circular photon orbit exists outside the BH.

\section{Photon Sphere, ISCO and Stability Analysis} \label{sec:PSISCOSA}
The requirements for circular orbits yield $V_{eff}(r)=E^2$ and $V'_{eff}(r)=0$. In case of null geodesics, at the radius of the circular orbits unstable photon spheres are formed. For timelike geodesics, at this position innermost circular orbits (ISCO) are formed, the position at which a test particle can stably orbit a BH in a perfect circle. At position $r > r_{isco}$, tiny perturbations applied to a circular orbit yield stable bounded radial oscillations, allowing the particle to safely recover its equilibrium and at $r < r_{isco}$, infinitesimal perturbations completely disrupt the orbital balance, making a circular trajectory impossible and forcing the particle to rapidly plunge inward. 
\subsection{Photon Sphere}
Circular null geodesics satisfy conditions 
\begin{equation}
    \frac{1}{b_c^2}=\frac{E^2}{L^2}=\frac{f(r_c)}{r_c^2},  \quad \quad 2 f(r_c)=r_c f'(r_c).
\end{equation}
where $r_c$ is the radius of the circular null geodesic, location of the photon sphere. In this region, the photons can move in unstable circular orbits under the influence of gravity. For a Schwarzschild black hole surrounded by an anisotropic fluid, the spacetime geometry is modified by the presence of dark energy like matter, which alters the location and properties of the photon sphere \cite{mishra2019understanding, gan2021photon}. In the present case, the photon sphere equation becomes 
\begin{equation}\label{eq:phsphere}
    \frac{6M}{r_{ph}} + \frac{3(\omega+1)N}{r_{ph}^{3\omega+1}} - 2 = 0.
\end{equation}
From Eqn. (\ref{eq:phsphere}) it is clear that the radius of photon sphere depends on the BH mass ($M$), the state parameter ($\omega$), and the normalization constant ($N$).
The photon sphere equation can be verified with the Schwarzschild photon sphere location as $r_{ph}=3M$ in the absence of the state parameter and normalization constant, i.e. $\omega=N=0$. Consequently, the photon sphere equations for different values of $\omega$ are
\begin{itemize}
    \item for $\omega=-1$,   \quad \quad   $r_{ph} = 3M$
    \item for $\omega=-2/3$, \quad  $r_{ph} = \frac{1 \pm \sqrt{1 - 6MN}}{N}$ 
    \item for $\omega=-1/3$, \quad   $r_{ph} = \frac{3M}{1 - N}$
    \item for $\omega= 0$,   \quad \quad \quad   $r_{ph} = 3(M+\frac{1}{2}N)$
    \item for $\omega=1/3$,  \quad \quad  $r_{ph} = \frac{3M \pm \sqrt{9M^2 + 8N}}{2}$
    \item for $\omega=1$,    \quad \quad \quad $r_{ph}^4-3Mr_{ph}^3-3 N=0 $
\end{itemize}
From the above photon sphere equations it is clear that the location of photon sphere is changing with different values of the state parameter. The radius of the photon sphere of a Schwarzschild black hole surrounded by an anisotropic fluid depends on the parameter $N$ (the value of this parameter is determined from Table \ref{table:N}), except $\omega=-1$ for which the radius of photon sphere is independent of $N$, indicates dark matter has no contribution in the photon sphere region.

\begin{figure}[htbp]
	\centering
	\subfigure[]{\includegraphics[scale=0.45]{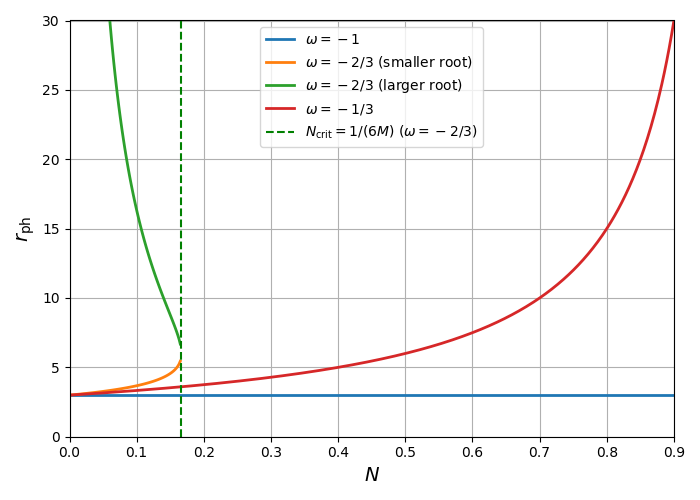}\label{fig:rphq}}
    \subfigure[]{\includegraphics[scale=0.45]{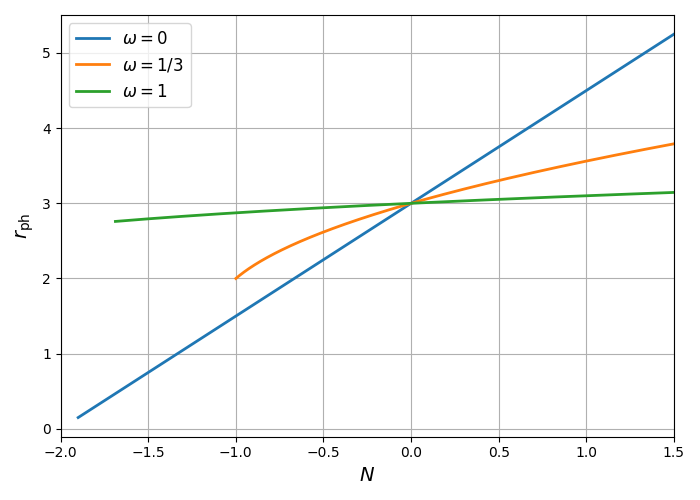}\label{fig:rphnq}}
	\caption {Variation of Photon sphere $r_{ph}$ with $N$, for  $\omega = -1,-2/3, -1/3$ in the upper panel and for  $\omega = 0, 1/3, 1$ in the lower panel. Here, we considered $M=1$.}\label{fig:rph}
\end{figure}

Fig. \ref{fig:rph} illustrates the variation of the photon sphere with $r$ for different values of the state parameter. In the upper panel, photon sphere is plotted for $\omega = -1,-2/3, -1/3$, and in the lower panel, for $\omega = 0,1/3, 1$,. It is clearly observed that all curves intersect at $N=0$, where the photon sphere radius reduces to the Schwarzschild case $r_{\rm ph}=3M$, which confirms that the Schwarzschild geometry is recovered in the absence of the surrounding-field contribution.
In Fig. \ref{fig:rphq}, for $\omega=-1/3$ and  $-2/3$, the photon-sphere radius increases monotonically with $N$, indicating that the surrounding field shifts the unstable circular null orbit towards larger radial distances. In case of $\omega=-2/3$, only the smaller positive root is physically admissible, since it lies within the static region bounded by the event and cosmological horizons, whereas the larger root lies beyond the cosmological horizon. In contrast, for $\omega=-1$, corresponding to the Schwarzschild–de Sitter spacetime which is independent of $N$, the photon-sphere radius remains constant at $r_{ph}=3M$.
In Fig. \ref{fig:rphnq},  for $\omega=0$, the photon sphere exhibits a linear dependence on $N$ but for $\omega=1/3$ and  $1$, the curves exhibit a nonlinear dependence on the surrounding-field parameter. Consequently, an increase in positive $N$ shifts the photon sphere towards larger radial distances, whereas negative values of $N$ lead to an inward shift. In the case of $\omega=1$, the radius of the photon sphere is obtained from the quartic equation. The resulting behaviour is nonlinear and differs from the other two cases. For negative $N$, the photon sphere equation may admit multiple positive real solutions, but the outer solution only considered which are lying in the static region.

\subsection{Innermost Circular Orbits (ISCO)}
ISCO is the smallest possible radius at which a test particle can stably orbit a BH in a perfect circle. At $r > r_{ISCO} $, tiny perturbations applied to a circular orbit yield stable, bounded radial oscillations, allowing the particle to safely recover its equilibrium. But at $r < r_{ISCO} $, the orbital balance completely disrupts and forces the particle to rapidly plunge
inward. 
The circular orbits conditions $V_{eff}(r)=E^2$ ($\dot{r}=0$) and $V'_{eff}=0$ ($\ddot{r}=0$) produce the angular momentum ($L$) and energy ($E$) for timelike geodesics as
\begin{equation}\label{eq:LandE}
    L^2 = \frac{r^3 f'(r)}{2f(r) - rf'(r)},  \quad \quad E^2 = \frac{2 f^2(r)}{2f(r) - rf'(r)}.
\end{equation}
Also,  the second derivative of the effective potential of the timelike geodesic is
\begin{equation}\label{eq:ddveff}
    (V_{eff}^T )''= f''(r_{c}) \left( 1+ \frac{L^2}{r_{c}^2}\right) - 4 f'(r_{c}) \frac{L^2}{r_{c}^3}+6 f(r_{c}) \frac{L^2} {r_{c}^2}.
\end{equation}
The ISCO equation can be solved by the inequality condition $\partial_{rr} V_{eff} \geq 0 $ \cite{PhysRevD.103.104070}.
Thus, with the value of $L^2$ from Eq. (\ref{eq:LandE}), the ISCO radius equation yields as
\begin{equation}\label{eq:ISCO}
   r_c f(r_{c})f''(r_{c})  - 2 r_c (f'(r_{c}))^2  + 3 f(r_{c}) f'(r_{c}) \geq 0.
\end{equation}
The radius of ISCO will be altered for a Schwarzschild BH surrounded by an anisotropic fluid as
\begin{equation}
    \begin{aligned}
    12M^2r_c^{6\omega} - 2Mr_c^{6\omega+1}
-6MNr_c^{3\omega}\left(3\omega^2-4\omega-2\right)\\
+Nr_c^{3\omega+1}
\left(9\omega^2-1\right)
+3N^2(3\omega+1)(\omega+1)\geq 0.
    \end{aligned}
\end{equation}
Evidently, depending on the value of the state parameter $\omega$, the ISCO equations for different values of $\omega$ are

\begin{figure}[htbp]
	\centering
	\subfigure[]{\includegraphics[scale=0.45]{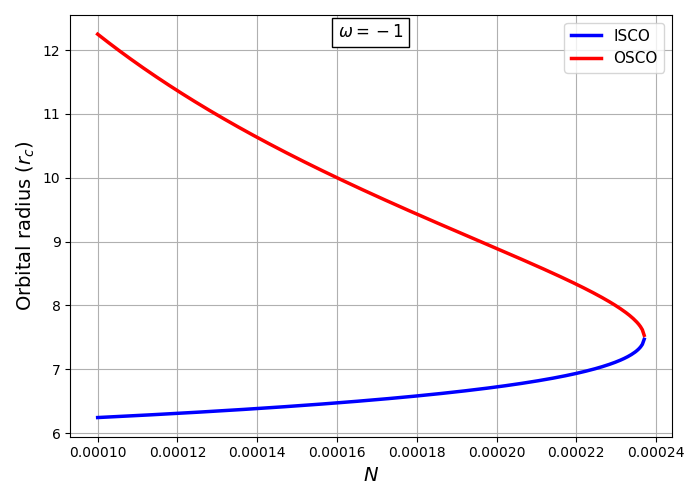}\label{fig:osco1}}
    \subfigure[]{\includegraphics[scale=0.45]{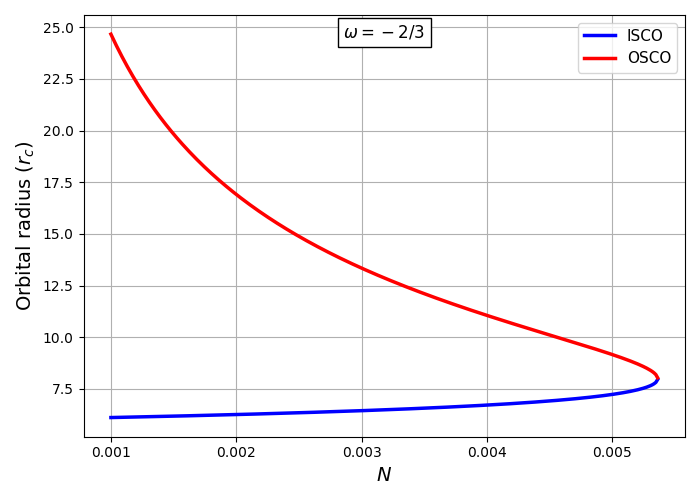}\label{fig:osco2}}
	\caption {Variation of ISCO radius ($r_{\rm ISCO}$) and OSCO radius ($r_{\rm OSCO}$) with $N$, for $\omega = -1$ (upper panel) and $\omega = -2/3$ (lower panel) with $M=1$.}\label{fig:riscoosco}
\end{figure}

\begin{itemize}
    \item for $\omega=-1$,   \quad   $4Nr_c^4 - 15MNr_c^3 - Mr_c + 6M^2 = 0$
    \item for $\omega=-2/3$, \\\\   $N^2r_c^4 - 3Nr_c^3 + 12MNr_c^2 + 2M r_c - 12M^2 = 0 $ 
    \item for $\omega=-1/3$, \quad   $r_{ISCO} = \frac{6M}{1-N}$
    \item for $\omega= 0$,   \quad\quad\quad $r_{ISCO}=6M+3N $
    \item for $\omega=1/3$,  \quad $Mr_c^3 - 6M^2r_c^2 - 9MNr_c - 4N^2 = 0$
    \item for $\omega=1$, \\\\ $Mr_c^7 - 6M^2r_c^6 - 4Nr_c^4 - 9MNr_c^3 - 12N^2 = 0 $ 
\end{itemize}
Therefore, the ISCO radius for different spacetimes, depending on the value of the state parameter $\omega$,  can be obtained by solving the above equations for the physically admissible positive root satisfying the circular-orbit conditions.
\begin{figure}[htbp]
	\centering
	\subfigure[]{\includegraphics[scale=0.45]{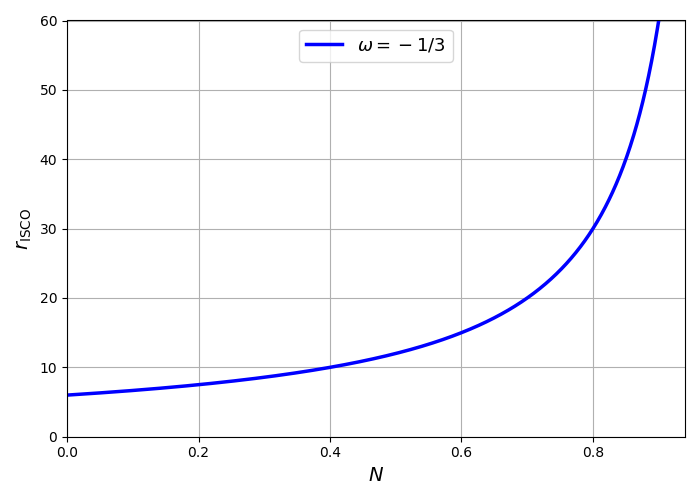}\label{fig:riscoq}}
    \subfigure[]{\includegraphics[scale=0.45]{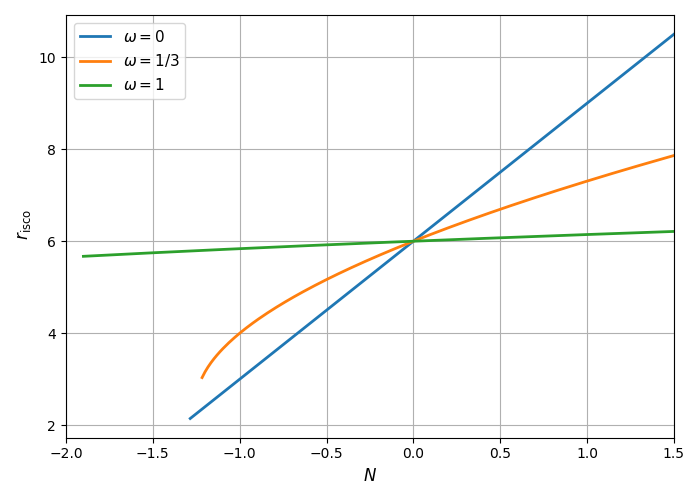}\label{fig:risconq}}
	\caption {Variation of inner most stable circular orbit (ISCO) radius $r_{ISCO}$ with $N$, for  $\omega = -1/3$ in the upper panel and for  $\omega = 0, 1/3, 1$ in the lower panel. Here, we considered $M=1$.}\label{fig:risco}
\end{figure}
Fig. \ref{fig:risco} illustrates the variation of the ISCO radius with $r$ for different values of equation of state parameter. In the upper panel, ISCO radius is plotted only for $\omega = -1/3$, and in the lower panel, for $\omega = 0, 1/3, 1$. It is observed that the radius of ISCO shifted while increasing $N$ and lead to an inward shift for negative value of $N$. From Fig. \ref{fig:risconq} it is clear that the ISCO radius of Schwarzschild BH can be recovered in absence of $N$.
From table \ref{table:ISCO}, it is seen that for $\omega=-\frac{2}{3}$ and $-1$, there are no physically accessible $r_{\rm ISCO}$ according to the consideration of $N$ values, where the available roots are out of the horizon ($r>r_h$) but do not lie in the static spacetime region ($f(r)<0$). But for other values of $\omega$, ISCO radius ($r_{\rm ISCO}$) increases with $N$. Referring to table \ref{tab:isco_osco} and Fig. \ref{fig:riscoosco}, we have the range of $N$ to obtain the physically accessible $r_{\rm ISCO}$ and $r_{\rm OSCO}$ (outermost stable circular orbit)  for $\omega=-\frac{2}{3}$ and $-1$. This table also gives the value of $N$ for which $r_{\rm ISCO}$ and $r_{\rm OSCO}$ are equal, which defines the ultimate environmental cutoff for a black hole. The OSCO provides a  numerical boundary separating a black hole that can actively interact with orbiting matter and form accretion structures from a black hole that has completely entered a macroscopically unstable, purely repulsive cosmic phase. Thus, from Fig. \ref{fig:riscoosco} it is  observed that while increasing the parameter $N$ (within the limiting range) the radius of ISCO increases and the radius of OSCO decreases non-linearly. At the critical limit of $N$, $r_{\rm isco}$ = $r_{\rm osco}$ and the stable circular orbit $(r_c)$ lies between ISCO and OSCO ($r_{isco}< r_c< r_{osco}$).
\begin{table}[htbp]
\centering
\caption{ISCO and OSCO radii for $\omega=-1$ and $-\frac{2}{3}$ with different values of the normalization parameter $N$ by considering ($M=1$).}
\centering
\begin{tabular}{|c|c|c|c|}
\hline
$\omega$ & $N$ & $r_{\rm ISCO}$ & $r_{\rm OSCO}$ \\ \hline
\multirow{3}{*}{$-1$}
& $0.000200$ & $6.722372$ & $8.891726$ \\ 
& $0.000237$ & $7.469022$ & $7.531528$ \\ 
& $\mathbf{0.000237037}$ &
$\mathbf{7.500000}$ & $\mathbf{7.500000}$ \\ \hline
\multirow{3}{*}{$-\frac{2}{3}$}
& $0.004000$ & $6.721292$ & $11.056933$ \\ 
& $0.005000$ & $7.237822$ & $9.162848$ \\ 
& $\mathbf{0.005361652}$ &
$\mathbf{8.000000}$ & $\mathbf{8.000000}$ \\ \hline
\end{tabular}\label{tab:isco_osco}
\end{table}

For fields characterized by $\omega=0$ (dust), $\omega = 1/3$ (radiation), and $\omega=1$ (stiff matter), an outermost stable circular orbit ($r_{\rm OSCO}$) is physically impossible. As these configurations gives an attractive gravitational addition rather than an outward cosmic push or drop off too quickly near the core, as in the stiff matter case, circular trajectories remain permanently stable all the way out to infinity, once they cross the inner ISCO boundary. Also, at $\omega = -1/3$, no outermost stable circular orbit can ever form as the quintessence field does not grow with $r$ at large distances for this specific equation of state and fails to provide any outward cosmic acceleration. Consequently, once an orbiting test particle passes beyond the inner gravitational stability barrier ($r > r_{\rm ISCO}$), its circular trajectory remains permanently stable up to infinity, provided that the normalization parameter satisfies the physical admissibility constraint $N < 1$. Radial layout of the stable and unstable domains governing timelike circular paths between the horizons ($r_h$ and $r_c$) is given below.

\begin{table}[htbp]
    \centering
    \begin{tabular}{|c|c|}
    \hline
        Inner Unstable Corridor &  $r_h < r < r_{\rm ISCO}$ \\
         Stable Orbital Window &  $r_{\rm ISCO} < r < r_{\rm OSCO}$  \\
         outermost Unstable Branch &  $ r_{\rm OSCO}<r<r_c$ \\
         \hline   
    \end{tabular}
\end{table}

\subsection{Stability Analysis using Lyapunov Exponent}
The stability or instability of the circular orbit can be determined using the Lyapunov exponent. 
The stability analysis can be investigated using the Jacobian matrix or the evolution matrix. 
Reformulating Eq. (\ref{eq:ddotr}), we get the non-linear differential equation in a two-dimensional phase-space ($r - p$), given by
\begin{equation}\label{eq:eom}
    \dot{r} = p_r, \quad \dot{p_r}=-\frac{1}{2} V_{eff}'(r)
\end{equation}
By linearizing the equation of motion (\ref{eq:eom}) about an orbit of constant radius $r=r_0$, the evolution matrix is expressed as
\begin{align}
    K &= \begin{pmatrix}
0 & \partial_r \dot{p_r} \\
\partial_p \dot{r}& 0
\end{pmatrix}\\
&= \begin{pmatrix}
0 & -\frac{1}{2} V_{eff}''(r) \\
1 & 0
\end{pmatrix}
\end{align}
The eigen values of the evolution matrix $\lambda$ along the circular orbits are called the principal Lyapunov exponent, expressed as 
\begin{equation}
    \lambda_t = \pm \sqrt{-\frac{V_{eff}''(r)}{2 }}.
\end{equation}
To maintain the conservation of the volume of phase space, the Lyapunov exponent is always in $\pm$ pairs. But in our calculations $\pm$ sign will be dropped, only positive Lyapunov exponent is to be considered. The conditions of stability and instability (unstable)  of the circular orbits are shown in table \ref{table:stability}. Also, a circular orbit is stable when $V_{eff}''(r) ~>~ 0$  and unstable when $V_{eff}''(r) ~<~ 0$.

Following the phase-space analysis, the coordinate time Lyapunov exponent is defined as \cite{PhysRevD.79.064016, pradhan2016stability, Shakir2026-ef}
\begin{equation}\label{eq:lyapexp}
    \lambda_t = \pm \sqrt{-\frac{V_{eff}''(r)}{2 \dot{t}^2}}.
\end{equation}
\begin{table}[htbp]
    \centering
    \caption{Classification of Stability using Lyapunov's linear stability analysis} \label{table:stability}
    \begin{tabular}{|c|c|}
    \hline
       Eigen values of Evolution Matrix  & Stability  \\
    \hline
       $\lambda_1 ~\leq ~\lambda_2 < 0$  &   Stable  \\
       $\lambda_1 ~\geq ~\lambda_2 > 0$  &   Unstable  \\
       $\lambda_1 < ~ 0  ~> \lambda_2 $ &   Unstable  \\
       $\lambda_1 ~=~\lambda_2 ~=~ 0 $  &   marginally stable \\
       $\lambda_{1,2} = a~\pm~ib$ with $a < 0$  &   Stable  \\
       $\lambda_{1,2} = a~\pm~ib$ with $a > 0$  &   Unstable  \\
       $\lambda_{1,2} = a~\pm~ib$ with $a = 0$  &   Stable  \\
    \hline
    \end{tabular}
\end{table}

The stability of the photon sphere is determined by using the above Eq. (\ref{eq:lyapexp}) at the radius of the photon sphere; where the second derivative of the effective potential of the null geodesic is   
\begin{equation}
    (V_{eff}^N )''= \left( f''(r_{c}) - 4 \frac{f'(r_{c})}{r_{c}}+6 \frac{f(r_{c})} {r_{c}^2}\right) \frac{L^2}{r_{c}^2},
\end{equation}
which gives the coordinate time Lyapunov exponent of the null geodesic as
\begin{equation}\label{eq:lambda_Null}
    \lambda_t^{N} = \pm \sqrt{- \frac{f(r_c)}{2 r_c^2} \bigg( r_c^2 f''(r_c) - 2 f(r_c)\bigg)}. 
\end{equation}
\begin{figure}[htbp]
	\centering
\includegraphics[width=\linewidth]{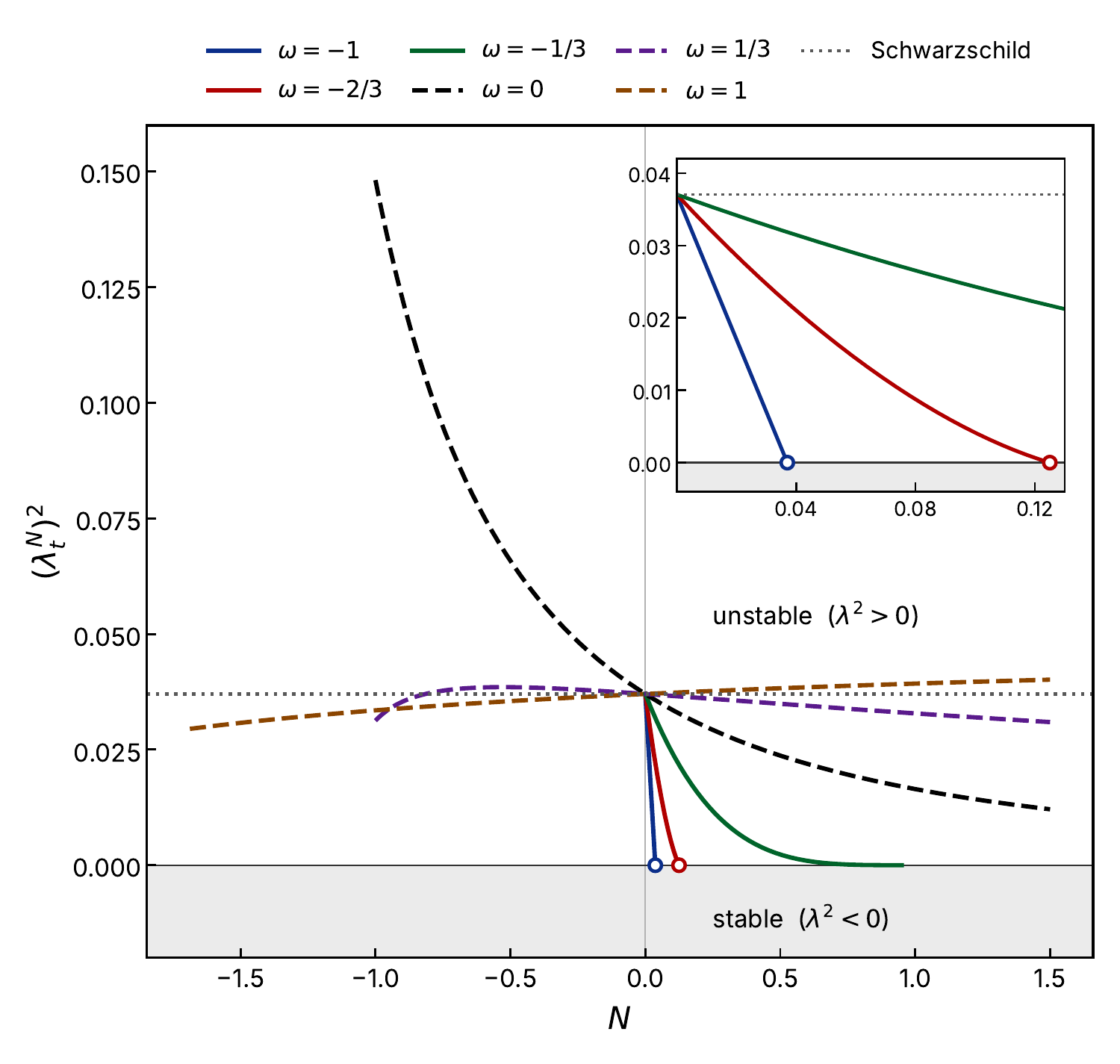}
	\caption{Square of the coordinate-time Lyapunov exponent, $(\lambda_t^N)^2$, for circular null geodesics at the photon sphere as a function of $N$, for different values of $\omega$ with $M=1$. The dotted line marks the Schwarzschild value $1/27$.}\label{fig:lyapunovnull}
\end{figure}
Thus, the Lyapunov exponent associated with a circular null orbit for a Schwarzschild black hole surrounded by an anisotropic fluid is
\begin{equation}\label{eq:lambda_general}
(\lambda_t^N)^2 = \frac{1}{2r_{\rm ph}^2} \left(1-\frac{2M}{r_{\rm ph}} -\frac{N}{r_{\rm ph}^{3\omega+1}}
\right) \left[ 2 + \frac{9\omega(\omega+1)N} {r_{\rm ph}^{3\omega+1}} \right].
\end{equation}
Thus, depending on the equation of state parameter ($\omega$), the Lynapunov exponent due to null geodesics depends on the normalization constant $N$ only for unit mass ($M=1$).
Table \ref{table:null_lyapunov} shows the photon sphere radius, the value of the positive Lynapunov exponent, and the nature of the circular orbit corresponding to different $N$ and $\omega$. The table clearly indicates that the radius of the photon sphere increases with $N$, irrespective of $\omega$, except $\omega=-1$ for which the photon sphere radius is independent of $N$.
\begin{table}[h]
\centering
\caption{Circular null orbit radii, Lyapunov exponents, and stability analysis for different values of $\omega$ and $N$, with $M=1$.}
\label{table:null_lyapunov}
\setlength{\tabcolsep}{7pt}
\renewcommand{\arraystretch}{1.3}
\begin{tabular}{|c|c|c|c|c|}
\hline
$\omega$ & $N$ & $r_c=r_{ph}$ & $ (\lambda_t^N)^2$ & Nature \\
\hline
\multirow{3}{*}{$-1$}
& 0.01 & 3.00000 & 0.02703704 & Unstable \\
& 0.02 & 3.00000 & 0.01703704 & Unstable \\
& 0.03 & 3.00000 & 0.00703704 & Unstable \\
\hline
\multirow{3}{*}{$-\frac{2}{3}$}
& 0.08 & 3.48612  & 0.00874650 & Unstable \\
& 0.09 & 3.57519  & 0.00630489 & Unstable \\
& 0.10 & 3.67544  & 0.00413418 & Unstable \\
\hline
\multirow{3}{*}{$-\frac{1}{3}$}
& 0.3 & 4.28571 & 0.00889259 & Unstable \\
& 0.5 & 6.00000 & 0.00231481 & Unstable \\
& 0.9 & 30.00000 & 0.00000370 & Unstable \\
\hline
\multirow{3}{*}{$0$}
& -1.5 & 0.75000 & 0.592592593 & Unstable \\
& -1.0 & 1.50000 & 0.148148148 & Unstable \\
& 1.0 & 4.50000 & 0.016460905 & Unstable \\
\hline
\multirow{3}{*}{$\frac{1}{3}$}
& -1.0 & 2.00000 & 0.031250000 & Unstable \\
& -0.8 & 2.30623 & 0.037227823 & Unstable \\
& 1.0 & 3.56155 & 0.032820125 & Unstable \\
\hline
\multirow{3}{*}{$1$}
& -1.6875 & 2.75893 & 0.029489385 & Unstable \\
& -1.0 & 2.87357 & 0.033497842 & Unstable \\
& 1.0 & 3.10064 & 0.039282820 & Unstable \\
\hline
\end{tabular}
\end{table}

Now with Eq. (\ref{eq:ddveff}) the coordinate time Lyapunov exponent of the timelike geodesic as
\begin{eqnarray}
    \lambda_t^{T} = \pm \sqrt{-\frac{1}{2r_c} \bigg[ r_c f(r_c) f''(r_c) - 2 r_c (f'(r_c))^2
     +  3  f(r_c) f'(r_c)\bigg]}\nonumber\\
\end{eqnarray}
Correspondingly, the Lyapunov exponent associated with a circular timelike orbit becomes
\begin{eqnarray}
(\lambda_t^T)^2
=&\frac{1}{2r_c^{6\omega+4}}
\Big[12M^2r_c^{6\omega} - 2Mr_c^{6\omega+1}\nonumber\\
&+6MN r_c^{3\omega} (-3\omega^2+4\omega+2)\nonumber\\
&   +3N^2(3\omega^2+4\omega+1)\nonumber\\
&+Nr_c^{3\omega+1}(9\omega^2-1) \Big]
\end{eqnarray}
\begin{figure}[htbp]
	\centering
\includegraphics[width=\linewidth]{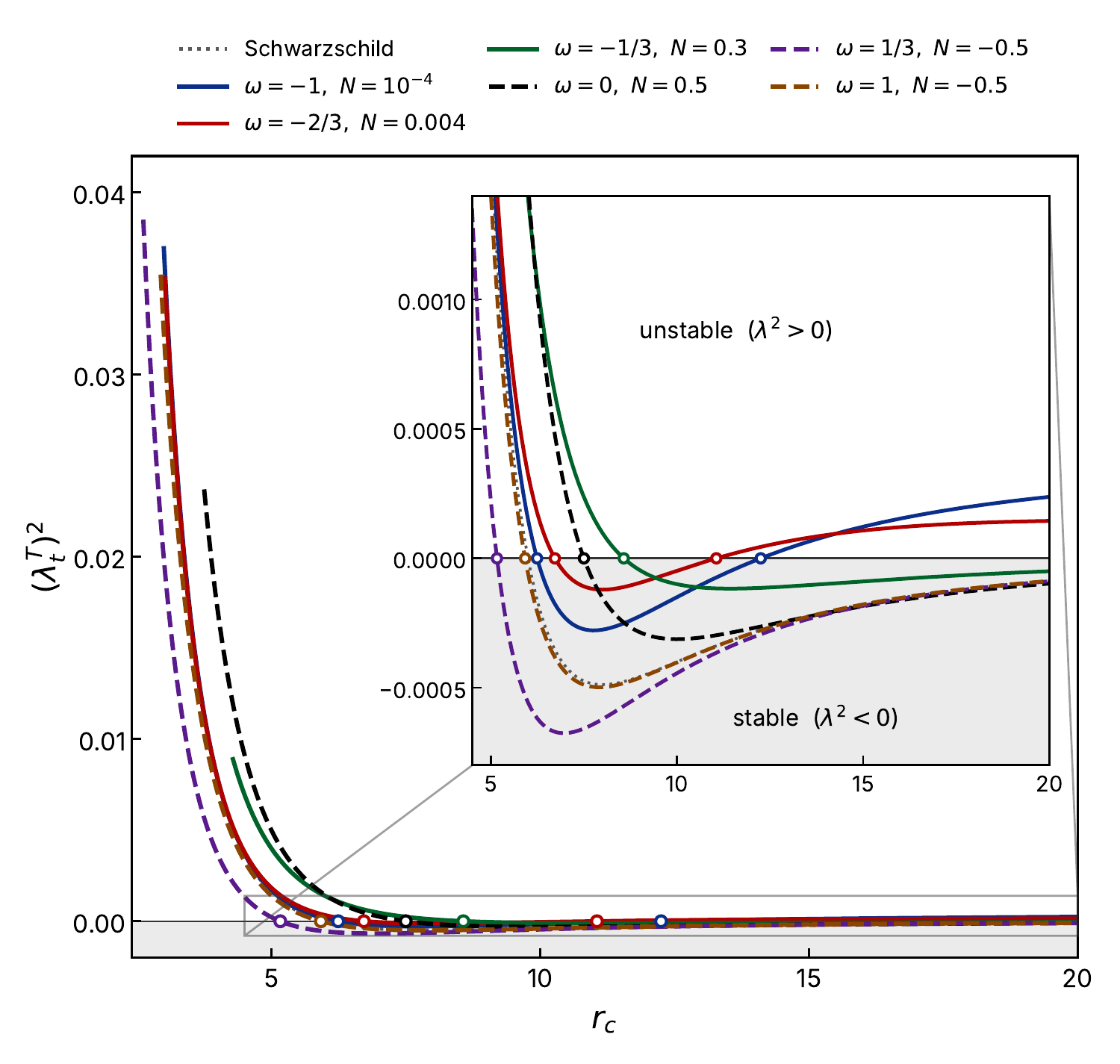}
	\caption{Square of the coordinate-time Lyapunov exponent $(\lambda_t^T)^2$, along the sequence of circular timelike orbits as a function of the orbit radius $r_c$ for different sets of the parameters $\omega$ and $N$ with $M=1$.}\label{fig:lyapunovtimelike}
\end{figure}
\begin{table*}[htbp]
\centering
\setlength{\tabcolsep}{4pt}
\renewcommand{\arraystretch}{1.25}
\caption{ISCO radius, radial positions corresponding to the maxima and  second derivatives corresponding to the maximum of the effective potential ($V''_{\max}$) for different values of $\omega$, $N$, and $L$, with $M=1$. Only physically admissible extrema in the static region are retained.}
\begin{tabular}{|c|c|c|cc|cc|cc|cc|}
\hline
\multirow{2}{*}{$\omega$} &
\multirow{2}{*}{$N$} &
\multirow{2}{*}{$r_{\rm ISCO}$} &
\multicolumn{2}{c|}{$L=2$} &
\multicolumn{2}{c|}{$L=4$} &
\multicolumn{2}{c|}{$L=5$} &
\multicolumn{2}{c|}{$L=10$} \\
& & & $r$ & $V_{max}''$ & $r$ & $V_{max}''$ & $r$ & $V_{max}''$ & $r$ & $V_{max}''$ \\   \hline
\multirow{3}{*} {-1} & $0.01$ & -- & 4.17922 & -0.07143 & 3.43854 & -0.23052 & 3.27843 & -0.41930 &  3.06693 & -2.22190 \\ 
& $0.02$ & -- & 3.47852 & -0.15961 & 3.21631 & -0.37881 & 3.14894 & -0.58042  & 3.04048 & -2.39795 \\
& $0.03$ & -- & 3.15147 & -0.25331 & 3.07533 & -0.52023 & 3.05421 & -0.73421 & 3.01609 &  -2.5710 \\ \hline
\multirow{3}{*} {-2/3} & $0.01$ & -- & 12.46008 & -0.00143 & 4.00000 & -0.06750 & 3.52956 & -0.23116 & 3.14333 & -1.86821\\
& $0.08$ & -- & 4.62618 & -0.03977 & 4.00000 & -0.10250 & 3.82897 & -0.17368 & 3.57317 & -0.87807 \\
& $0.09$ & -- & 4.44149 & -0.04774 & 4.00000 & -0.10750 & 3.86906 & -0.16941 & 3.65418 & -0.76933\\ \hline
\multirow{3}{*} {-1/3} & $0.3$ & 8.57143 & -- & -- &  -- & -- & 7.50000 & -0.00158 & 4.58619 & -0.27499 \\
& $0.5$ & 12.00000 & -- & -- & -- & -- & -- & -- & 6.97224 & -0.03052 \\
& $0.9$ & 60.00000 & -- & -- & -- & -- & -- & -- & -- & -- \\ \hline
\multirow{4}{*} {0} & $-1.5$ & 1.50000 & 0.78890 & -13.06429 & 0.75900 & -61.26123 & 0.75571 & -99.19827 & 0.75141 & -410.86545 \\
& $-1$ & 3.00000 & 2.00000 & -0.25000 & 1.57779 & -4.65436 & 1.54792 & -8.16990 & 1.51142 & -37.74614 \\
& $0.5$ & 7.50000 & -- & -- & -- & -- & 5.00000 & -0.02400 & 3.94449 & -0.31294 \\
& $1.5$ & 10.50000 & -- & -- & -- & -- & -- & -- & 5.84861 & -0.04604 \\ \hline
\multirow{3}{*} {1/3} & $-1$ & 4.00000  & -- & -- & 2.34315 & -0.19949 & 2.19224 & -0.42857 & 2.04168 & -4.81186 \\
& $-0.5$ & 5.16646  & -- & -- & 3.25170 & -0.05436 & 2.95017 & -0.14710 & 2.68651 & -2.42848 \\
& $0.5$ & 6.69451 & -- & -- & 4.79596 & -0.00392 & 3.95861 & -0.02766 & 3.42639 & -0.99971 \\ \hline
\multirow{3}{*} {1} & $-1.6875$ & 5.71268  & -- & -- & 3.69699 & -0.01633 & 3.22200 & -0.06317 & 2.85098 &  -1.25055 \\
& $-0.5$ & 5.92135 & -- & -- & 3.91908 & -0.00840 & 3.41806 & -0.04149 & 3.03518 &  -1.04018 \\
& $0.5$ & 6.07434  & -- & -- & 4.07570 & -0.00445 & 3.54852 & -0.02890  & 3.15040 & -0.91008 \\ \hline
\end{tabular}\label{table:Vmax}
\end{table*}

Fig.~\ref{fig:lyapunovnull} shows that circular photon orbits remain unstable for all considered values of $\omega$ and $N$, although the strength of the instability depends significantly on the surrounding matter distribution (see also Table~V). For $\omega=-1$, $-2/3$ and $-1/3$, increasing $N$ weakens the instability by lowering and flattening the effective potential barrier. The photon-sphere radius remains unchanged for $\omega=-1$, shifts outwards for $\omega=-2/3$, and grows without bound as $N\to1$ for $\omega=-1/3$. For $\omega=-1$ and $-2/3$ the instability vanishes at the extremal value of $N$, where the event and cosmological horizons merge. For $\omega=0$, positive $N$ acts as an additional mass contribution, shifting the photon sphere outwards and reducing its instability. For $\omega=1/3$, where negative $N$ plays the role of an electric charge with $Q^2=-N$, the instability depends non-monotonically on $N$: it slightly exceeds the Schwarzschild value for moderately negative $N$ and decreases towards the extremal limit and for positive $N$. In contrast, for $\omega=1$, positive $N$ sharpens the potential barrier and enhances the photon-sphere instability.

The stability of circular orbits of massive particles, illustrated in Fig.~\ref{fig:lyapunovtimelike}, depends even more strongly on the surrounding matter distribution, since stable orbits can disappear altogether. The squared Lyapunov exponent $(\lambda_t^T)^2$ is positive inside the ISCO and changes sign exactly at the ISCO and, where it exists, at the OSCO, so that these radii delimit the stable orbital region, while the depth of the negative region indicates the strength of radial stability. The ISCO radii, locations of the maxima of the effective potential, and corresponding second derivatives for different parameter values are summarized in Table~\ref{table:Vmax}. For $\omega=-1$ and $-2/3$, the repulsive contribution dominates at large distances, so that for $N$ below a critical value the stable region is finite and bounded by the ISCO and OSCO; between the OSCO and the static radius, $(\lambda_t^T)^2$ becomes positive again. Increasing $N$ shifts the ISCO outwards and the OSCO inwards and makes the stable region shallower, progressively narrowing it until both boundaries coincide at the critical value of $N$. Beyond this value, stable circular orbits cease to exist. For $\omega=-1/3$, no OSCO develops and stable circular orbits extend to infinity, although increasing $N$ shifts the ISCO outwards and makes the stable region shallower. For $\omega=0$, the geometry is that of a Schwarzschild BHwith mass $M+N/2$, so positive $N$ moves the ISCO outwards and reduces the depth of the stable region. Finally, for $\omega=1/3$ and $\omega=1$, the modifications primarily affect the inner boundary of the stable region. For $\omega=1/3$, negative $N$ shifts the ISCO inwards and deepens the stable region, whereas for $\omega=1$ the ISCO undergoes only a slight displacement.

\section{Conclusion} \label{sec:conclusion}

In this work, we have investigated the geodesic structure and stability of circular orbits around a Schwarzschild BH surrounded by a Kiselev anisotropic fluid, characterized by the EoS parameter $\omega$ and the normalization parameter $N$. We have considered six representative values of $\omega$, namely $\omega=-1,-2/3,-1/3,0,1/3,$ and $1$, and examined how the surrounding matter distribution modifies the horizon structure, effective potentials, photon sphere, ISCO, OSCO, and the stability of null and timelike circular orbits. The main findings of our analysis are summarized as follows:

\begin{enumerate}[label=(\roman*)]
    \item The horizon structure is strongly influenced by the normalization parameter $N$ and the EoS parameter $\omega$. Depending on their values, the spacetime may possess a single event horizon, an event horizon accompanied by a cosmological horizon, or an inner and an outer horizon. Extremal configurations arise when two horizons merge. In particular, the extremal limits for $\omega=-1$ and $-2/3$ correspond to the merging of the event and cosmological horizons, whereas those for $\omega=1/3$ and $1$ correspond to the merging of the inner and outer horizons for negative $N$. For $\omega=-1/3$ and $0$, the spacetime possesses a single horizon throughout the admissible parameter range. These results indicate that the surrounding field modifies the causal structure of the spacetime and determines the region in which physically meaningful circular motion can occur.

    \item TThe effective potential analysis shows that increasing the angular momentum of a massive test particle strengthens the centrifugal barrier, while the surrounding anisotropic fluid modifies its height and radial location. For $\omega=-1$ and $-2/3$, no physically admissible minimum exists for the selected values of $N$, which exceed the corresponding critical values. Thus, stable circular orbits are absent. In contrast, for $\omega=-1/3$, $0$, $1/3$, and $1$, the potential can exhibit both a maximum and a minimum, corresponding to unstable and stable circular orbits, respectively. For massless particles, the potential has a single maximum at the photon sphere throughout the physically accessible region, bounded by the event horizon and, where present, the cosmological horizon. Although additional minima appear for $\omega=1/3$ and $1$ with $N<0$, they lie either on the degenerate horizon in the extremal case or between the inner and outer horizons, where $f(r)<0$. Hence, they do not represent physical circular photon orbits, and no stable circular photon orbit exists outside the BH.

    \item The photon-sphere radius exhibits a distinct dependence on the surrounding matter distribution. For $\omega=-1$, it remains fixed at $r_{\rm ph}=3M$, independently of $N$. For $\omega=-2/3$ and $-1/3$, it increases with $N$, while for $\omega=0$ it varies linearly with $N$. The cases $\omega=1/3$ and $1$ exhibit a nonlinear dependence on the normalization parameter. The Lyapunov analysis confirms that all physically admissible photon spheres remain unstable within the parameter space considered. The strength of this instability decreases with increasing $N$ for $\omega=-1$, $-2/3$, $-1/3$, and $0$, and vanishes at extremality for $\omega=-1$ and $-2/3$. For $\omega=1/3$, it exhibits non-monotonic behaviour, whereas for $\omega=1$ it increases with $N$. Notably, for $\omega=-1$, the instability changes with $N$ even though the photon-sphere radius remains unchanged.

    \item Our analysis reveals the existence of an OSCO for $\omega=-1$ and $-2/3$, in addition to the conventional ISCO, consistent with earlier studies~\cite{Hussain:2016tth}. For $N$ below a critical value $N_c$, stable circular orbits are confined to a finite radial interval between the ISCO and OSCO. Increasing $N$ shifts the ISCO outwards and the OSCO inwards, progressively narrowing the stable orbital region. At the critical values, $N_c\approx0.000237$ for $\omega=-1$ and $N_c\approx0.005361$ for $\omega=-2/3$, the two radii coincide at $r=7.5$ and $r=8$, respectively. Beyond these critical values, no stable circular orbits exist. These results are summarized in Table~\ref{tab:isco_osco}. For $\omega=-1/3$, $0$, $1/3$, and $1$, no physically admissible OSCO develops within the parameter ranges considered, and stable circular orbits extend outwards from the ISCO.

    \item The coordinate-time Lyapunov exponent provides an independent dynamical characterization of the stability of circular timelike orbits, consistent with the effective potential analysis. We have derived a closed-form expression for $(\lambda_t^T)^2$ for arbitrary $\omega$ and demonstrated that its sign changes precisely at the marginally stable circular-orbit boundaries, namely the ISCO and, wherever it exists, the OSCO. Positive values indicate unstable circular orbits, whereas negative values correspond to stable circular motion. The vanishing of the exponent therefore identifies the transition between the two regimes.

    \item In the appropriate limits, our results recover several well-known BH geometries. Setting $N=0$ yields the Schwarzschild solution, while $\omega=-1$, $0$, and $1/3$ (with $N=-Q^2$ in the last case) reproduce the Schwarzschild--de Sitter, mass-shifted Schwarzschild, and Reissner--Nordstr\"om geometries, respectively. These results show that different surrounding matter distributions can produce substantially different orbital properties around a BH of the same mass, and they provide a systematic picture of how an anisotropic fluid modifies the geodesic dynamics and stability of the Schwarzschild BH. They may therefore have implications for observable phenomena such as BH shadows, thin accretion disk and quasi-periodic oscillations.
\end{enumerate}

A natural extension of this work would be to compare the Lyapunov exponents obtained here with the surface gravity $\kappa=f'(r_h)/2$ of the event horizon to investigate the chaos bound $\lambda\leq\kappa$ proposed by Maldacena, Shenker, and Stanford~\cite{Maldacena:2015waa,Hashimoto:2016dfz}. Such a comparison would be particularly interesting near the extremal limits, where $\kappa\to0$, whereas $\lambda_t^N$ remains finite for $\omega=1/3$ and $1$.
\section*{Acknowledgments}
The authors, SR and PD express sincere and deep gratitude to the Department of Physics, NITA, for providing the necessary research environment to complete this work. One of the authors, HN would like to thank IUCAA, Pune, for the support under its associateship program, where a part of this work was done. The author, HN acknowledges financial support from the Anusandhan National Research Foundation (ANRF), through the Science and Engineering Research Board (SERB) Core Research Grant (Grant No. CRG/2023/008980).
\section*{Data Availability Statement }
{This manuscript has no associated data. All results presented in this work are based on analytical calculations and numerical computations described in the manuscript.}
\bibliographystyle{ieeetr}
\bibliography{SBHkiselev}
\end{document}